\documentclass[%
 reprint,
 superscriptaddress,
 amsmath,amssymb,
 aps,
 prb,
]{revtex4-2}

\usepackage{graphicx}% Include figure files
\usepackage{dcolumn}% Align table columns on decimal point
\usepackage{bm}% bold math
\usepackage{hyperref}% add hypertext capabilities
\usepackage[caption=false]{subfig}
\usepackage[export]{adjustbox}
\usepackage{dblfloatfix}
\usepackage{braket}
\newcommand{\phantomsubfloat}[1]{% Allows referencing phantom subfigures, i.e. when (a), (b) are part of an image itself.
 {% apply caption setup only temporarily
  \captionsetup[subfloat]{farskip=0pt,captionskip=0pt}
  \captionsetup[subfigure]{labelformat=empty}
  \subfloat{#1}
 }%
}

\begin{document}

\preprint{APS/123-QED}

\title{Antidot superlattices in two-dimensional topological insulators}% Force line breaks with \\

\author{Haolin Huang}
\email{haolinhuang0@gmail.com}
\affiliation{School of Science, RMIT University, Melbourne 3000, Australia}

\author{Jackson S. Smith}
\affiliation{School of Science, RMIT University, Melbourne 3000, Australia}

\author{Fabio Taddei}
\affiliation{Istituto Nanoscienze -- CNR, NEST-SNS, Piazza San Silvestro 12, 56127 Pisa, Italy}

\author{Michele Governale}
\affiliation{School of Chemical and Physical Sciences and MacDiarmid Institute for Advanced Materials and Nanotechnology, Victoria University of Wellington, PO Box 600, Wellington 6140, New Zealand}

\author{Jared H. Cole}
\email{jared.cole@rmit.edu.au}
\affiliation{School of Science, RMIT University, Melbourne 3000, Australia}

\date{\today}

\begin{abstract}
We investigate the electronic properties of a two-dimensional topological insulator patterned with an array of circular holes.
The band structures of the topological insulator superlattices are calculated using the Bernevig-Hughes-Zhang model, discretized with the finite element method.
The band topology is studied using the Fukui-Hatsugai-Suzuki method, exploiting the fact that topological charge is conserved when a band gap closes and reopens.
We find that when the holes are close to each other and the edge states overlap significantly, the material loses its topological character and becomes a trivial insulator.
The results show that patterning topological insulators changes their properties, and this can be achieved with a feature size compatible with current lithographic techniques.
This shows the potential to incorporate patterned topological insulators in future electronics.
\end{abstract}

%\keywords{Suggested keywords}%Use showkeys class option if keyword display desired
\maketitle

%\tableofcontents

\section{\label{sec:intro}Introduction}

In 2005, \citeauthor{Kane2005} proposed that the spin-orbit interaction can facilitate dissipationless spin currents on the edge of graphene \cite{Kane2005}.
In contrast to the quantum Hall effect, this does not rely on an applied magnetic field.
This became known as the quantum spin Hall (QSH) effect.
Later, the QSH effect was realised in mercury telluride-cadmium telluride quantum wells, which exhibits a stronger spin-orbit interaction \cite{BHZ}.
Despite the presence of a band gap, the band structure of materials that exhibit the QSH effect is topologically distinct from that of an insulator \cite{Kane2005-Z2}, hence these materials are called topological insulators (TIs).
Aside from theoretical interest, the dissipationless spin currents also have applications in low-energy electronics and spintronics \cite{Konig2008}, and topological insulators have since became a popular area of research \cite{Weber2024}.

To make electronic devices, one needs to be able to manipulate the properties of materials.
In the semiconductor industry, this is achieved by doping the material.
However, the topological protection of the conducting states in TIs is robust against impurity scattering \cite{Asboth2016,Vaitkus2022}, limiting the effectiveness of doping in modifying their conduction properties.
As an alternative, since most of the physics of TIs happens on the boundary their properties can be altered by changing the boundary.
This can be achieved by patterning the material.

One way to pattern the material is to make a superlattice.
A superlattice is a lattice of material structures, which means a repeating pattern with periodicity much greater than that of the underlying crystal lattice.
In 1970, \citeauthor*{Esaki1970} predicted that superlattices would subdivide the Brillouin zone into minizones and create previously forbidden states in the band gap \cite{Esaki1970}.
This allows for band structure engineering, creating metamaterials with artificial properties.
Researchers can engineer features with sizes similar to the wavelength of electromagnetic radiation, enabling interactions with light that were previously thought impossible, like having a negative refractive index \cite{Shelby2001,Pendry2004}.
If the feature size decreases to the Fermi wavelength of electrons, interesting interactions with electrons become possible \cite{Song2018}.

The honeycomb lattice is particularly interesting.
A famous natural example is graphene as the first well-known 2D material.
Its electronic properties were predicted last century \cite{Wallace1947,Semenoff1984,DiVincenzo1984}, and it was successfully isolated from graphite at the beginning of this century \cite{Novoselov2004}.
It showed high electronic and heat conductivity, and has since became a focus of research \cite{Novoselov2012}.
This inspired researchers to investigate artificial structures of the same shape, which extended the parameter space beyond that of graphene \cite{Polini2013}.
These structures are called artificial graphene (AG), and include
Moir\'{e} lattices induced by twisted bilayer graphene \cite{Cao2018super,Cao2018insul},
gate-induced quantum dots on Si \cite{Chen2021},
triangular antidot lattices on GaAs \cite{Du2018,Krix2020,Krix2022,Wang2023,Wang2026},
Haldane lattice in GaAs heterostructures \cite{Cioni2026},
and organic molecules assembled on metal \cite{Wang2013}.

In this paper, we study the electronic behaviour in holey 2D TIs, which are 2D TIs with a regular array of holes fabricated to form a superlattice \cite{Tretiakov2011,Maier2017,Niyazov2023}.
Previous studies have considered the role of geometric confinement in modifying the properties of TIs \cite{Gioia2019,Governale2020,Governale2023}.
Here, we investigate how circular holes (antidots) affect the spin Hall conductance.
To achieve this, we calculate the spin Chern number of the periodic structure, as the spin Hall conductance is directly related to the accumulated spin Chern number in the occupied bands \cite{Thouless1982,Niu1985,Sheng2006}.
As fabricated holes are usually smooth and the shape can be arbitrary, the finite element method (FEM) is used in this study.

Previous work considered superlattices in mercury telluride quantum wells using a tight-binding method on a square grid to represent hexagonal holes \cite{Fu2018}.
Representing such a superlattice on a square grid requires a high-resolution tight-binding lattice to accurately represent the holes.
In our work, we mitigate this effect using a FEM approach.
In addition, this approach allows us to consider arbitrary hole shapes and, in particular, circular holes which are difficult to represent on a square grid.
Superlattices in graphene were previously studied using a tight-binding model focused on atomic-scale holes \cite{Kariyado2018,Jiang2026}.
For mesoscopic devices where the characteristic length scales are much larger than the underlying lattice spacing, our work is more efficient than an atomistic description of the underlying lattice structure.
Topological effects were also studied in photonic crystals \cite{Wang2020,Yu2021}, but photonic crystals do not have a valence band, making the calculation of band topology much simpler.
Our work provides a method to account for the numerous valence bands that shows up in electronic structure calculations, for which the band topology calculation would otherwise be intractable. By applying this method, we calculate the spin Chern number of the main gap to determine whether the system is trivial or topological. We obtain a phase diagram in terms of the hole period and the ratio of the hole diameter to the period. We find a phase transition from the QSH state to the trivial insulating state driven by the overlap of edge states localized around the antidots.

The paper is organized as follows.
In Section~\ref{sec:methods}, we detail the model and methods used for our calculations: the model Hamiltonian and the system of study are specified in Subsection~\ref{sec:methods-bhz}, the method for computing band topology is explained in Subsection~\ref{sec:methods-chern}, and the topological argument to simplify such process for electronic band structure is introduced in Subsection~\ref{sec:methods-bulk-to-model}.
In Section~\ref{sec:results}, we present a topological phase diagram for different periods and sizes of the holes, with the study of the edge states localized around the holes in Subsection~\ref{sec:results-localized} and the investigation of the topological phase transition in Subsection~\ref{sec:results-transition}.

\section{\label{sec:methods}Model and methods}

\subsection{\label{sec:methods-bhz}The BHZ Hamiltonian}

In 2006, \citeauthor*{BHZ} (BHZ) proposed an effective continuum Hamiltonian for the conduction and valence bands closest to the Fermi level for telluride quantum wells \cite{BHZ,Qi2011}:
\begin{align}
    H(\mathbf{k}) &= \varepsilon(\mathbf{k}) \sigma_0 \tau_0 + M(\mathbf{k}) \sigma_0 \tau_z + A(k_x \sigma_z \tau_x - k_y \sigma_0 \tau_y) \nonumber\\
    &= \varepsilon(\mathbf{k}) + \begin{pmatrix}
        M(\mathbf{k}) & A k_+ & 0 & 0 \\
        A k_- & -M(\mathbf{k}) & 0 & 0 \\
        0 & 0 & M(\mathbf{k}) & -A k_- \\
        0 & 0 & -A k_+ & -M(\mathbf{k})
    \end{pmatrix} ,
    \label{eq:bhz}
\end{align}
where
\begin{align}
    k_\pm &= k_x \pm ik_y , \nonumber \\
    \mathbf{k} &= [k_x, k_y] , \nonumber\\
    \varepsilon(\mathbf{k}) &= C - D k^2 , \nonumber\\
    M(\mathbf{k}) &= M - B k^2 ,
\end{align}
and $\sigma_i/\tau_i$ are the Pauli matrices in the spin/orbital spaces respectively.
In the equations above,
$A$ is the spin-orbit coupling,
$B$ is the Newtonian mass,
$M$ is the Dirac mass, and
$D$ breaks the electron-hole symmetry.
The parameter $C$ shifts the energy by a constant, thus fixing the Fermi energy.
These parameters depend on the quantum well geometry, and can be obtained from $\mathbf{k} \cdot \mathbf{p}$ theory calculations \cite{Beugeling2025}.
It is also possible to include block off-diagonal bulk inversion asymmetry (BIA) and structural inversion asymmetry (SIA) terms that depends on $\mathbf{k}$ \cite{Liu2008}, but they are negligible and not considered in this work \cite{Qi2011}.
For large enough TIs such that the edge states are well-separated, the model has been solved exactly to give a linear dispersion:
\begin{equation}
    E_{k,\uparrow/\downarrow} = E_0 \pm \hbar v_0 k ,
    \label{eq:edge-energy}
\end{equation}
where the energy of Dirac point is $E_0 = C - MD/B$ and the state velocity is $v_0 = \sqrt{B^2 - D^2}A/(\hbar |B|)$ \cite{Zhou2008,Wada2011,Lunde2012}.
We set the energy scale to $|M|$ and the length scale to:
\begin{equation}
    r_0 = \frac{A}{|M|} ,
\end{equation}
which is roughly the decay length of the edge mode (see Appendix~\ref{app:analytic}).

\begin{figure}[!htb]
    \centering
    \includegraphics[width=\linewidth]{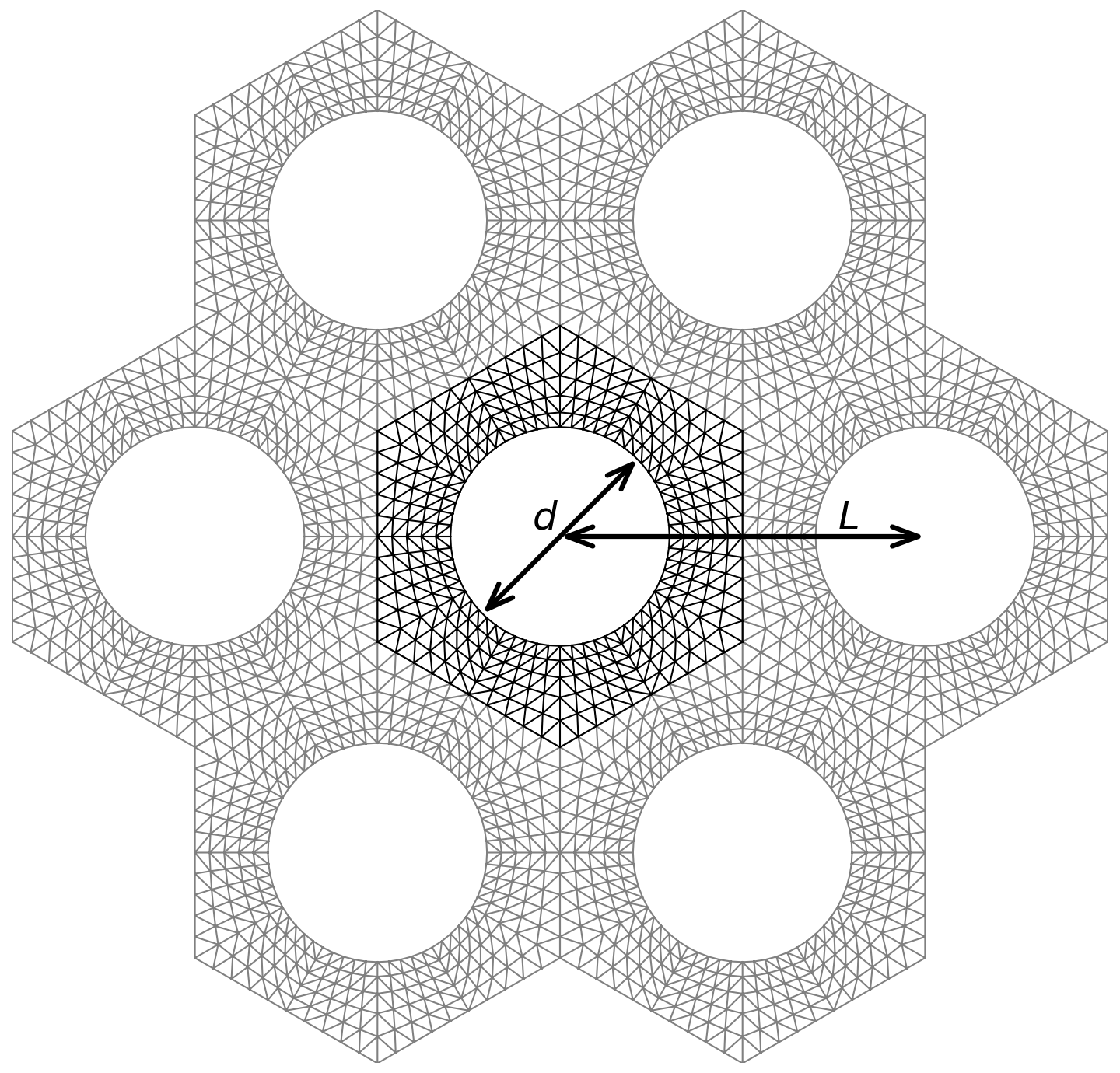}
    \caption{\label{fig:hex-mesh}
    Schematic of the meshing of a hole array with hexagonal symmetry, with $d$ and $L$ labeled.
    The gray meshes are copies of the black mesh across the periodic boundaries.
    Note that the schematic is only indicative and the calculations were performed on finer meshes.}
\end{figure}

In this work, we model 2D TIs with a lattice of circular holes forming antidots \cite{Tretiakov2011,Maier2017,Niyazov2023}.
The lattice of interest is a hexagonal lattice of period $L$.
Each hole has a diameter $d < L$.
Despite the TIs having an underlying crystal structure, the feature size and characteristic length are much larger than the lattice constant of the crystal, hence the geometry can be considered smooth and we use FEM to discretize the continuum model.
We employ Gmsh \cite{Geuzaine2009} for meshing the system and FEniCSx \cite{Baratta2023} to setup the Schr\"{o}dinger equation over the mesh.
See Fig.~\ref{fig:hex-mesh} for the schematic of the meshed system.

\begin{table}[!htb]
    \centering
    \caption{\label{tab:parameters}
    Parameters used in this study.}
    \begin{tabular}{lrl}
        \hline\hline
        $A$ & $345$ & meV$\cdot$nm \\
        $B$ & $-917$ & meV$\cdot$nm$^2$ \\
        $D$ & $-739$ & meV$\cdot$nm$^2$ \\
        $M$ & $-16.89$ & meV \\
        \hline\hline
    \end{tabular}
\end{table}

Table~\ref{tab:parameters} lists the parameters used in this study, which are for a 8~nm thick HgTe/Hg$_{0.32}$Cd$_{0.68}$Te quantum well, calculated by the Kdotpy Python package \cite{Beugeling2025}.
These amount to a length scale of $r_0 = 20.4$~nm.
We set $C = 0$, so the conduction and valence band edges are at $\pm |M|$.
The Dirac point is then at $E_0 = C - MD/B = 0.806 |M| = 13.6$~meV.

\subsection{\label{sec:methods-chern}Topological classification of the QSH state}

In 1982, \citeauthor*{Thouless1982} (TKNN) derived a formula to calculate quantized Hall conductance \cite{Thouless1982}:
\begin{align}
    \sigma_H = \frac{e^2}{h} \sum_m \frac{i}{2\pi} \int_\text{BZ} d^2 \mathbf{k} &  % break formula into 2 lines
    \left(\Braket{\frac{\partial u_m}{\partial k_1}|\frac{\partial u_m}{\partial k_2}}
    \right. \nonumber\\ & \left.  % break formula into 2 lines
    - \Braket{\frac{\partial u_m}{\partial k_2}|\frac{\partial u_m}{\partial k_1}}\right) ,
\end{align}
where $m$ is the band index, the sum runs over the filled bands, the integral is over the Brillouin zone, and $\ket{u_m(\mathbf{k})}$ is the eigenstate of band $m$ at wavevector $\mathbf{k}$.
The prefactor $e^2/h$ is the quantum of conductance, and the summand was shown to be an integer for each isolated band.
The integer was recognized as a topological invariant, called the TKNN invariant.
Specifically, if the band gap between two bands closes and reopens as the system parameters vary, the TKNN invariant of each band can change while the sum is preserved, i.e.\ the topological charge is exchanged between the bands \cite{Avron1983}.

Mathematically, the TKNN invariant of band $m$ is equal to its Chern number $c_m$, which is calculated as \cite{Simon1983}:
\begin{equation}
    c_m = \frac{1}{2\pi} \int_\text{BZ} d^2 \mathbf{k} \mathcal{F}_m ,
    \label{eq:chern}
\end{equation}
where $\mathcal{F}_m = \nabla \times i\braket{u_m(\mathbf{k})|\nabla|u_m(\mathbf{k})}$ is the Berry curvature of the band.
The quantized Hall conductance can then be succinctly expressed as
\begin{equation}
    \sigma_H = \frac{e^2}{h} \sum_{m=0}^{N} c_m .
\end{equation}

In 2005, \citeauthor*{Fukui2005} (FHS) provided an efficient method using lattice gauge theory to compute the Chern number of a band \cite{Fukui2005}.
The Brillouin zone is discretized into a lattice along the two reciprocal vectors $\mathbf{b}_1$ and $\mathbf{b}_2$.
The eigenstates $\ket{u_m(\mathbf{k})}$ for an isolated band $m$ are solved for each lattice point $\mathbf{k}$.
Instead of discretizing the differential operator in the expression of Berry curvature, a $U(1)$ link variable is defined:
\begin{equation}
    U_\mu(\mathbf{k}) = \frac{\braket{u_m(\mathbf{k})|u_m(\mathbf{k} + \mathbf{b}_\mu)}}{|\braket{u_m(\mathbf{k})|u_m(\mathbf{k} + \mathbf{b}_\mu)}|} ,
\end{equation}
where $\mu = 1, 2$, i.e.\ a complex phase factor with the same argument as $\braket{u_m(\mathbf{k})|u_m(\mathbf{k} + \mathbf{b}_\mu)}$ as long as it is non-zero.
The Berry curvature integrated on each plaquette can then be evaluated as
\begin{equation}
    \tilde{F}_m(\mathbf{k}) = \operatorname{Arg} U_1(\mathbf{k}) U_2(\mathbf{k} + \mathbf{b}_1) U_1(\mathbf{k} + \mathbf{b}_2)^{-1} U_2(\mathbf{k})^{-1} ,
\end{equation}
where $\operatorname{Arg}$ is the argument in the principle branch, i.e.\ the complex phase accumulated from link variables around the plaquette.
The Chern number can be evaluated by substituting this for $\mathcal{F}_m$ in Eq.~\ref{eq:chern}.
The result converges quickly to the real Chern number even for a coarse lattice, hence the efficiency.
This method can be extended to calculate the collective Chern number of multiple bands that are isolated from the rest, by replacing the link variable with
\begin{equation}
    U_\mu(\mathbf{k}) = \frac{\det\braket{u_m(\mathbf{k})|u_{m'}(\mathbf{k} + \mathbf{b}_\mu)}}{|\det\braket{u_m(\mathbf{k})|u_{m'}(\mathbf{k} + \mathbf{b}_\mu)}|} ,
\end{equation}
where $\braket{u_m(\mathbf{k})|u_{m'}(\mathbf{k} + \mathbf{b}_\mu)}$ is a matrix for row index $m$ and column index $m'$ running over these bands.
This is especially useful in the case of degeneracy, e.g.\ when multiple bands cross each other.

The Chern number described above counts the quantized conductance of the quantum Hall effect.
However, to describe the QSH effect, a different topological invariant is required.
In 2005, \citeauthor{Kane2005-Z2} showed that the QSH phase is related to a $\mathbb{Z}_2$ topological order \cite{Kane2005-Z2}.
In 2006, \citeauthor{Sheng2006} proposed that the spin Hall conductance is related to the spin Chern number \cite{Sheng2006}, which is now commonly defined as $c_s = (c_\uparrow - c_\downarrow)/2$, the half difference between the Chern numbers $c_{\uparrow/\downarrow}$ of the spin-up and spin-down bands.
These two classifications were later shown to be equivalent, with an odd spin Chern number signifying the QSH phase \cite{Prodan2009}.
The Chern number for bands of opposite spin must be opposite due to the time-reversal symmetry, i.e.\ $c_\uparrow + c_\downarrow = 0$, hence the spin Chern number can be expressed as $c_s = c_\uparrow$ \cite{Hung2014}.

In this study, we evaluate the spin Chern number by solving for eigenstates of the spin-up block of the BHZ Hamiltonian and calculate the Chern number using the FHS method.

To make the explanation clear, we need to label the bands and the gaps between them.
Hereinafter, we label the bulk band gap with the index $m = 0$, where the Fermi level lies.
The other gaps are labeled sequentially in increasing order of energy, so the next gap above gap $m$ will have index $m+1$.
Each band shares the same index as the gap directly above it, so the top valence band has index $m = 0$, and all bands are also labeled sequentially in increasing order of energy.

We use $c_m$ to denote the Chern number of band $m$ and $C_m$ to denote the accumulated Chern number up to the gap $m$.
$C_m$ is then the sum of Chern numbers of all bands below and including band $m$:
\begin{equation}
    C_m = \sum_{i \leq m} c_i = C_{m-1} + c_m .
    \label{eq:gap-chern}
\end{equation}
Then, an odd $C_0$ signifies the QSH phase and an even $C_0$ signifies the trivial phase.

\begin{widetext}
\begin{figure*}[!h]
    \centering
    \includegraphics[width=\linewidth]{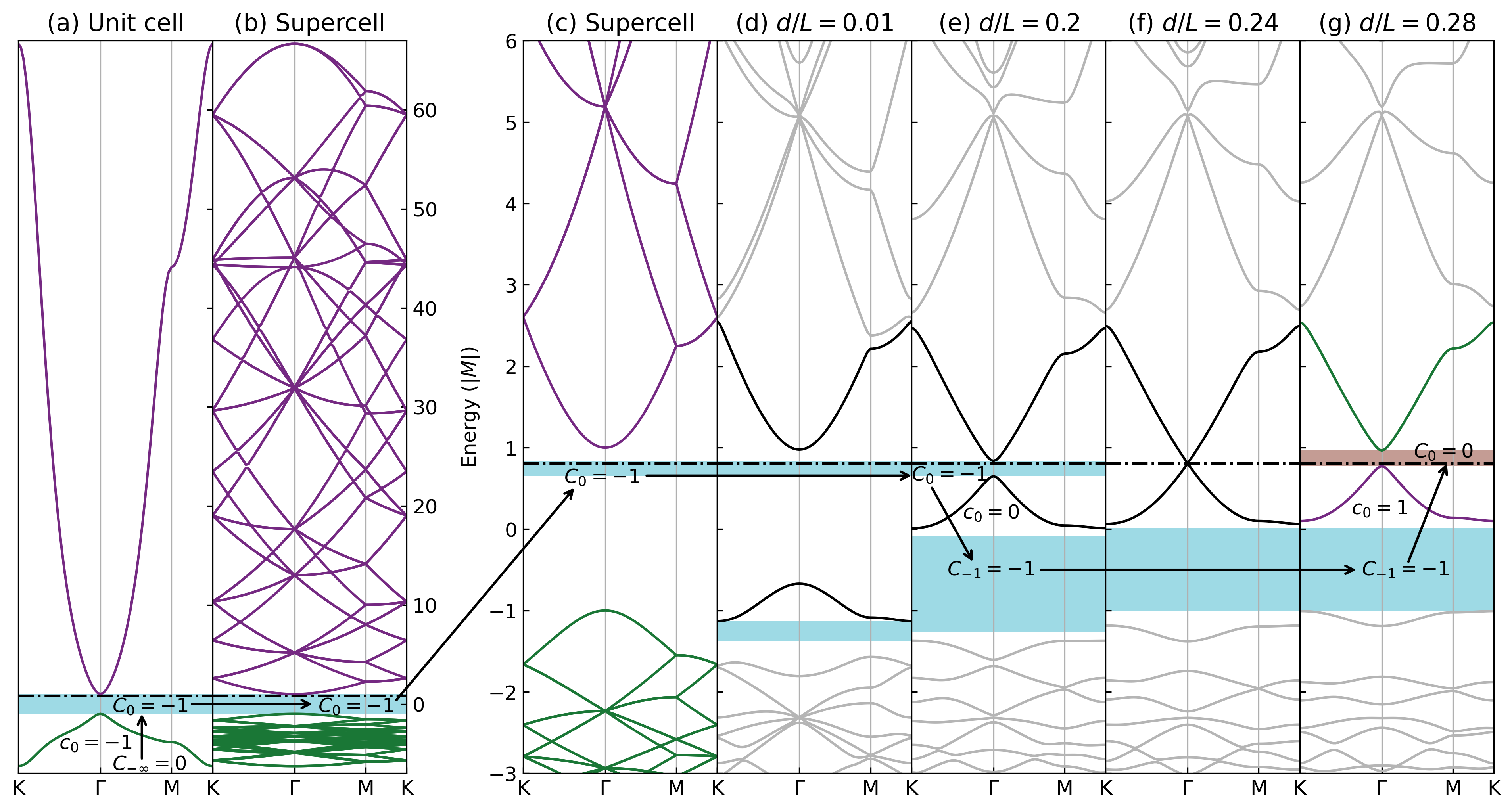}
    \phantomsubfloat{\label{fig:bulk-to-model-a}}
    \phantomsubfloat{\label{fig:bulk-to-model-b}}
    \phantomsubfloat{\label{fig:bulk-to-model-c}}
    \phantomsubfloat{\label{fig:bulk-to-model-d}}
    \phantomsubfloat{\label{fig:bulk-to-model-e}}
    \phantomsubfloat{\label{fig:bulk-to-model-f}}
    \phantomsubfloat{\label{fig:bulk-to-model-g}}
    \caption{\label{fig:bulk-to-model}
    This is an illustrative diagram for Section~\ref{sec:methods-bulk-to-model}, where the calculation of $C_0$ for each system is carried out following a procedure schematically indicated here by the black arrows.
    (a-c) The band structures of a system with no hole (effectively $d = 0$), computed with
    (a) a rhombic unit cell that is equivalent to a hexagonal unit cell of $L = 0.5 r_0$, and
    (b-c) a hexagonal supercell of $L = 2 r_0$, consisting of the unit cells described in (a).
    Band folding relative to (a) can be observed in (b). Panel (c) is a zoomed-in view from (b).
    (d-g) The band structures of a hexagonal supercell of $L = 2 r_0$ with a hole of radius $d$, with the relative $d/L$ ratio described in the title of each panel.
    In all panels, the dash-dotted lines indicate the energy of the Dirac point.
    The band directly above or below the bulk gap (whether or not folded) with a Chern number of $c = 1$, $c = 0$, and $c = -1$ is colored purple, black, and green, respectively.
    The light blue and light brown shading lie in a gap with accumulated Chern number of $C = -1$ and $C = 0$ respectively.}
\end{figure*}
\end{widetext}

\subsection{\label{sec:methods-bulk-to-model}Evaluating the Chern number}

Since the number of numerically solved occupied bands scales as the number of sample points in the discretization, computing the Chern number becomes intractable for a geometry with enough details.
Here, we propose an approach to simplify the computation.

A system with no holes is equivalent to bulk, where a unit cell with only one sample point and periodic boundary conditions suffices for the calculation of the band structure.
In this case, there is only one valence band, and its Chern number $c_0$ can be easily calculated with the FHS method.
$C_0$ is equal to $c_0$ as there is only one valence band.
This is illustrated in Fig.~\ref{fig:bulk-to-model-a}.

A supercell may be formed by making an array of unit cells.
This causes band folding, which creates multiple valence bands as shown in Fig.~\ref{fig:bulk-to-model-b}.
However, since this is merely changing the description of the same system, the band gap must be identical to before.
$C_0$ therefore must be the same.

When a tiny hole is introduced in each supercell, it perturbs the band structure slightly as shown in Fig.~\ref{fig:bulk-to-model-d}, which can be considered as a continuous deformation from the bulk band structure in Fig.~\ref{fig:bulk-to-model-c}.
Importantly, the band gap does not close, so $C_0$ stays the same.
The same argument applies as the hole size continuously increases as long as the band gap stays open.

As the hole size is increased further, the band gap closes and reopens, as shown in Fig.~\ref{fig:bulk-to-model-e} to \ref{fig:bulk-to-model-g}.
$c_0$ and $c_1$ may change, causing $C_0$ to change.
In this case, $c_0$ and $c_1$ need to be calculated before the band gap closes and after it reopens to confirm the change.
Since the two bands are isolated from all other bands, this is also easily achieved with the FHS method.
Specifically, $C_{-1}$ stays the same before the band gap closes and after it reopens, so the change in $C_0$ is the same as the change in $c_0$ as Eq.~\ref{eq:gap-chern} implies.
In this way, $C_0$ can be calculated for all geometries, without having to solve for all eigenstates in the occupied bands.

Since for the systems studied here, the band gap always closes at the $\Gamma$ point, a golden section search was used to find when the conduction band reaches a minimum (equivalently when the valence band reaches a maximum) at the $\Gamma$ point.

\begin{widetext}
\begin{figure*}[!h]
    \centering
    \includegraphics[width=0.9\linewidth]{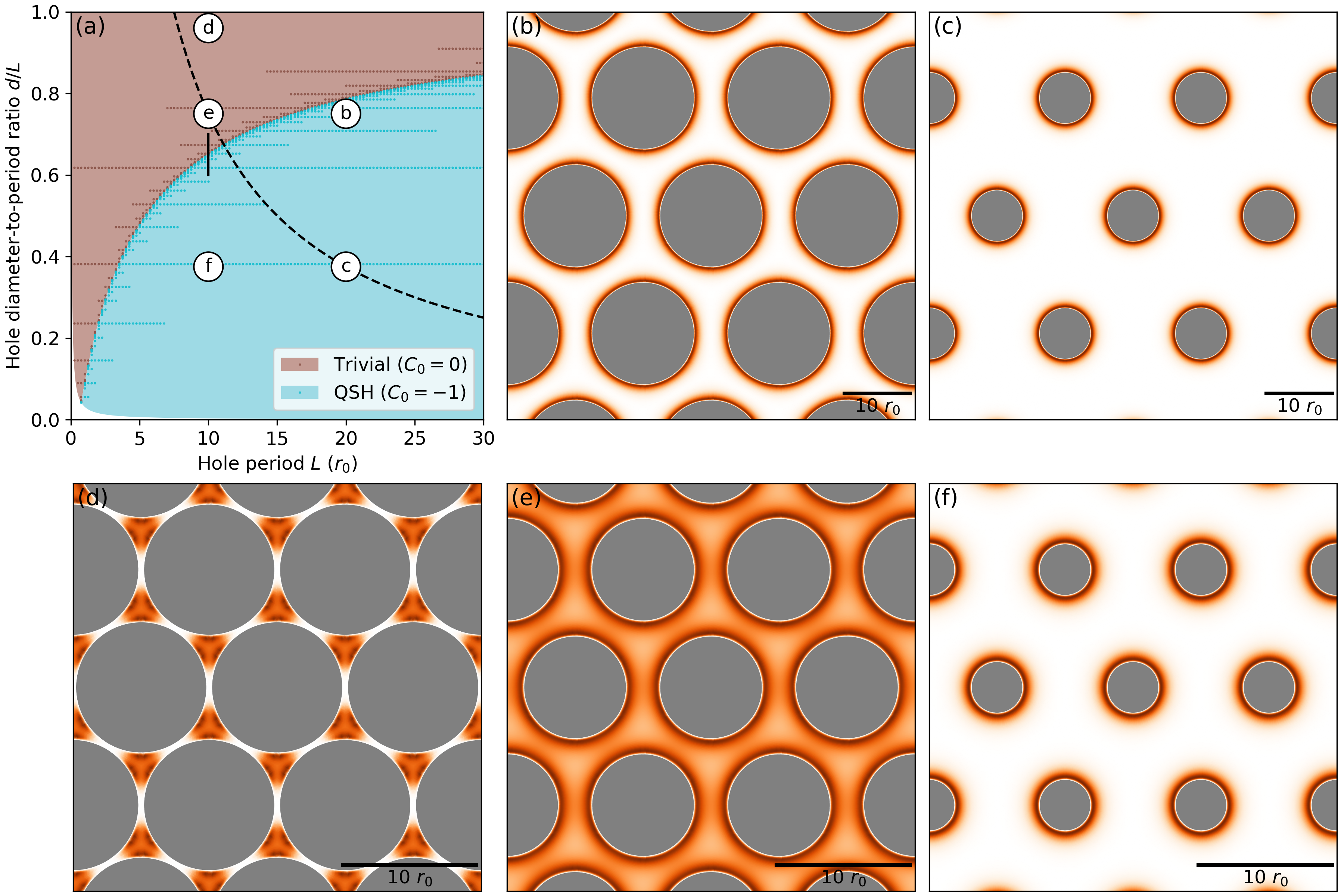}
    \phantomsubfloat{\label{fig:hex-phase}}
    \phantomsubfloat{\label{fig:hex-state-b}}
    \phantomsubfloat{\label{fig:hex-state-c}}
    \phantomsubfloat{\label{fig:hex-state-d}}
    \phantomsubfloat{\label{fig:hex-state-e}}
    \phantomsubfloat{\label{fig:hex-state-f}}
    \caption{\label{fig:hex-state}
    (a) A phase diagram for varying hole period and diameter.
    The dashed line is $d = 7.5 r_0$.
    The white shaded region is $d < 0.646 \text{nm} = 0.0316 r_0$.
    (b-f) Probability density for the states at the top of the valence band (at the $\Gamma$ point) for hole periods and diameters labeled in (a).
    Darker orange means higher probability.
    The gray shaded region is the holes.
    The scale bar is $10 r_0$.
    Only the hexagonal region around the hole in the center is used for simulation.
    The plots are created by repeating the region with respect to the periodic boundary conditions as in Fig.~\ref{fig:hex-mesh}.}
\end{figure*}
\end{widetext}

\section{\label{sec:results}Results}
In this section, we present the main results of this paper. We start by obtaining a phase diagram for varying hole period and diameter. We perform a golden section search to find the minimum (maximum) of the conduction (valence) band at the $\Gamma$ point and thus identify the value of $d/L$ where the band gap closes for every value of $L$. The boundary between the trivial phase and QSH phase can be traced out on the phase diagram by connecting the transition points.
The resulting phase diagram is shown in Fig.~\ref{fig:hex-phase}, with the dots marking the systems simulated during the search.

Fig.~\ref{fig:hex-state-b} to \ref{fig:hex-state-f} show the probability density of the highest energy occupied state ($|\Psi_\text{vbm}|^2$) for various systems marked in Fig.~\ref{fig:hex-phase}.
This state corresponds to an energy at the top of the valence band, which varies depending on the diameter and spacing of the holes.
They show that the state is localized around the holes when the system is in QSH phase (Fig.~\ref{fig:hex-state-b}, \ref{fig:hex-state-c}, and \ref{fig:hex-state-f}).

The transition from Fig.~\ref{fig:hex-state-f} to \ref{fig:hex-state-e} shows that the QSH phase can be broken by increasing the hole size.
The transition from Fig.~\ref{fig:hex-state-c} to \ref{fig:hex-state-e} shows that the QSH phase can be broken by bringing the holes closer to each other.
The transition from Fig.~\ref{fig:hex-state-b} to \ref{fig:hex-state-e} shows that the QSH phase can be broken by scaling the system down, i.e.\ decreasing $d$ and $L$ while keeping the ratio $d/L$ constant.

The state is delocalized in the spaces between the holes when the system is in the trivial insulating phase (Fig.~\ref{fig:hex-state-d} and \ref{fig:hex-state-e}).
In the case of Fig.~\ref{fig:hex-state-d}, the state is actually localized in the interstices left by the holes.
These results show that the QSH phase breaks when the holes are close to each other and the localized states overlap sufficiently.

\begin{figure}[!b]
    \centering
    \includegraphics[height=0.4\textheight]{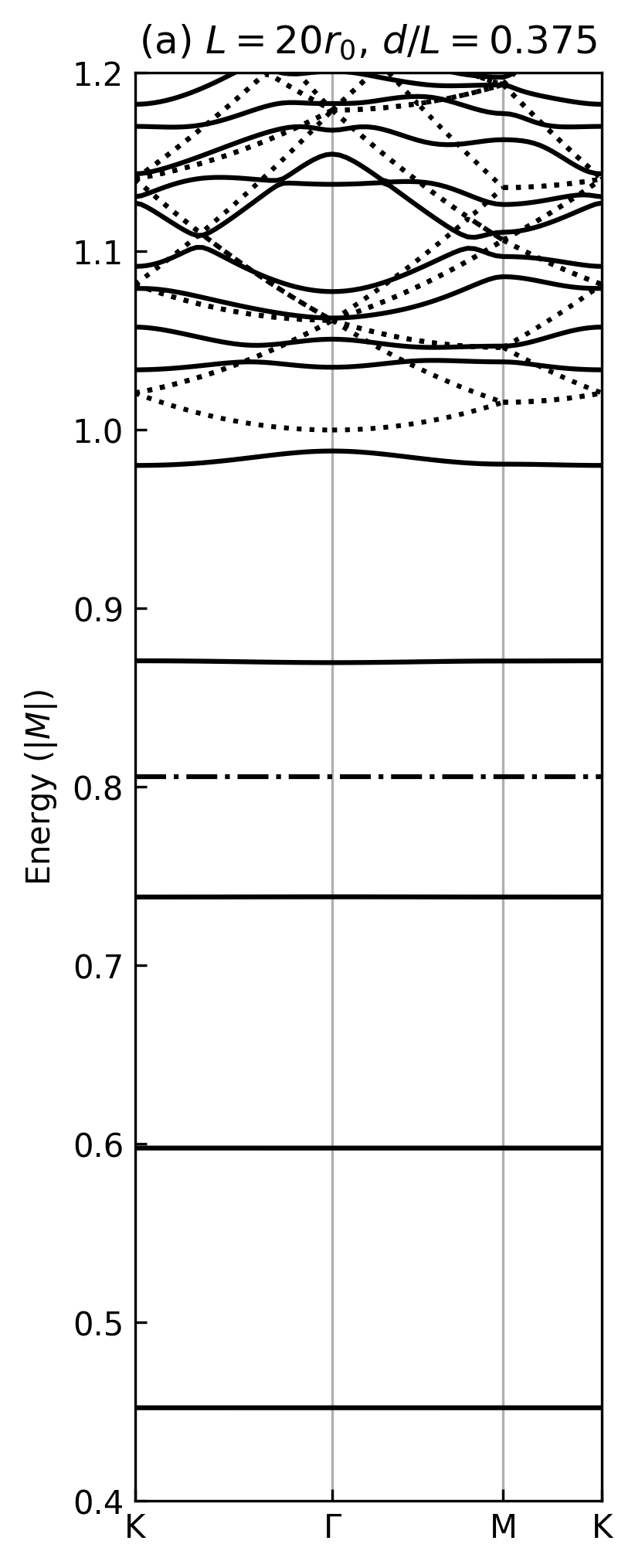}
    \includegraphics[height=0.4\textheight]{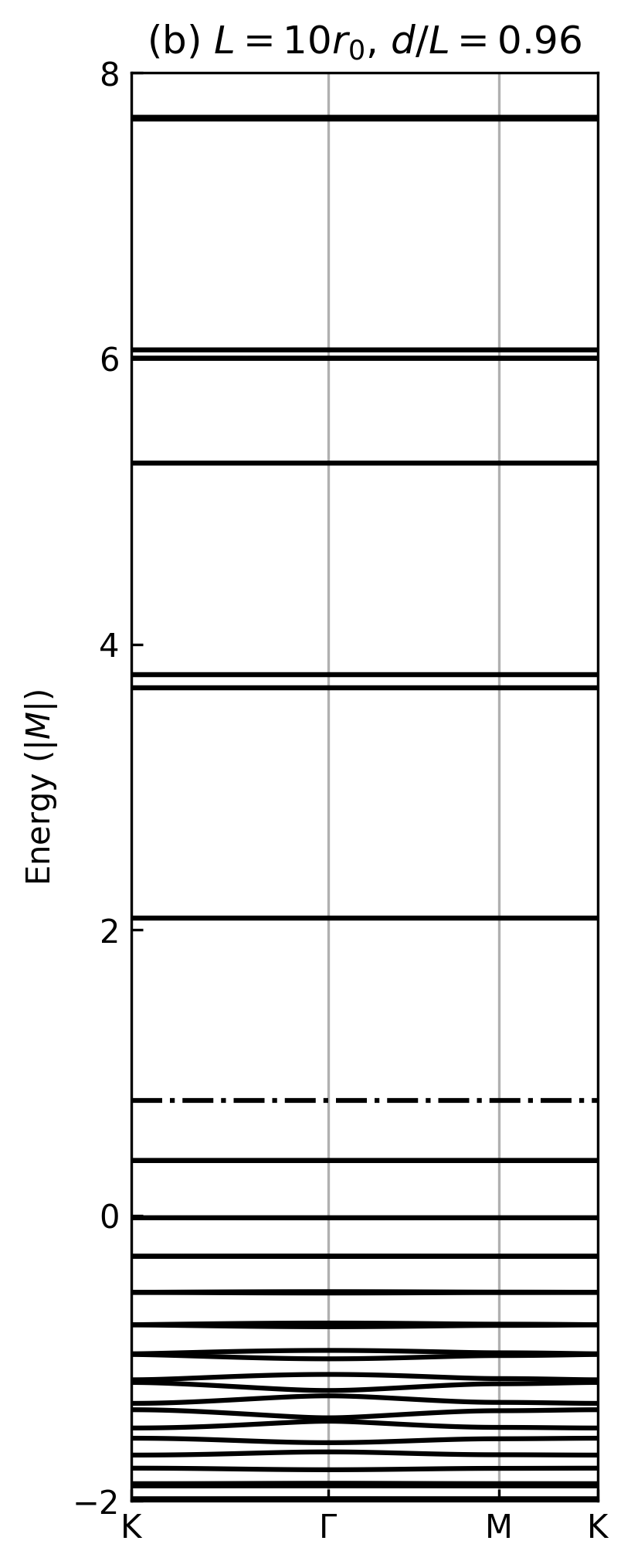}
    \phantomsubfloat{\label{fig:hex-bands-flat}}
    \phantomsubfloat{\label{fig:hex-bands-confined}}
    \caption{\label{fig:hex-bands}
    The band structures for (a) the system in Fig.~\ref{fig:hex-state-c} and (b) the system in Fig.~\ref{fig:hex-state-d}.
    The $L$ and $d/L$ values of the systems are labeled in the subfigure titles.
    The dash-dotted line is the energy of the Dirac point.
    The dotted lines in (a) is the band structure for a system with no holes.}
\end{figure}

Since the BHZ model is a continuum model, it breaks down when the feature size is comparable to the lattice constant of the crystal, where the system can no longer be considered as a continuum.
Therefore, we masked out the region of $d < 0.646 \text{nm} = 0.0316 r_0$ in Fig.~\ref{fig:hex-phase}, which corresponds to systems with holes smaller than one lattice constant of mercury telluride.

The right boundary of Fig.~\ref{fig:hex-phase} is $L = 30 r_0 = 612.8$~nm.
This means the features described in this study are on the order of hundreds of nanometers, which can be fabricated with current lithographic techniques \cite{Bake2023}.
This shows that it is feasible with existing technology to tune the topological property of TIs by patterning.

\begin{figure}[!t]
    \centering
    \includegraphics[width=\linewidth]{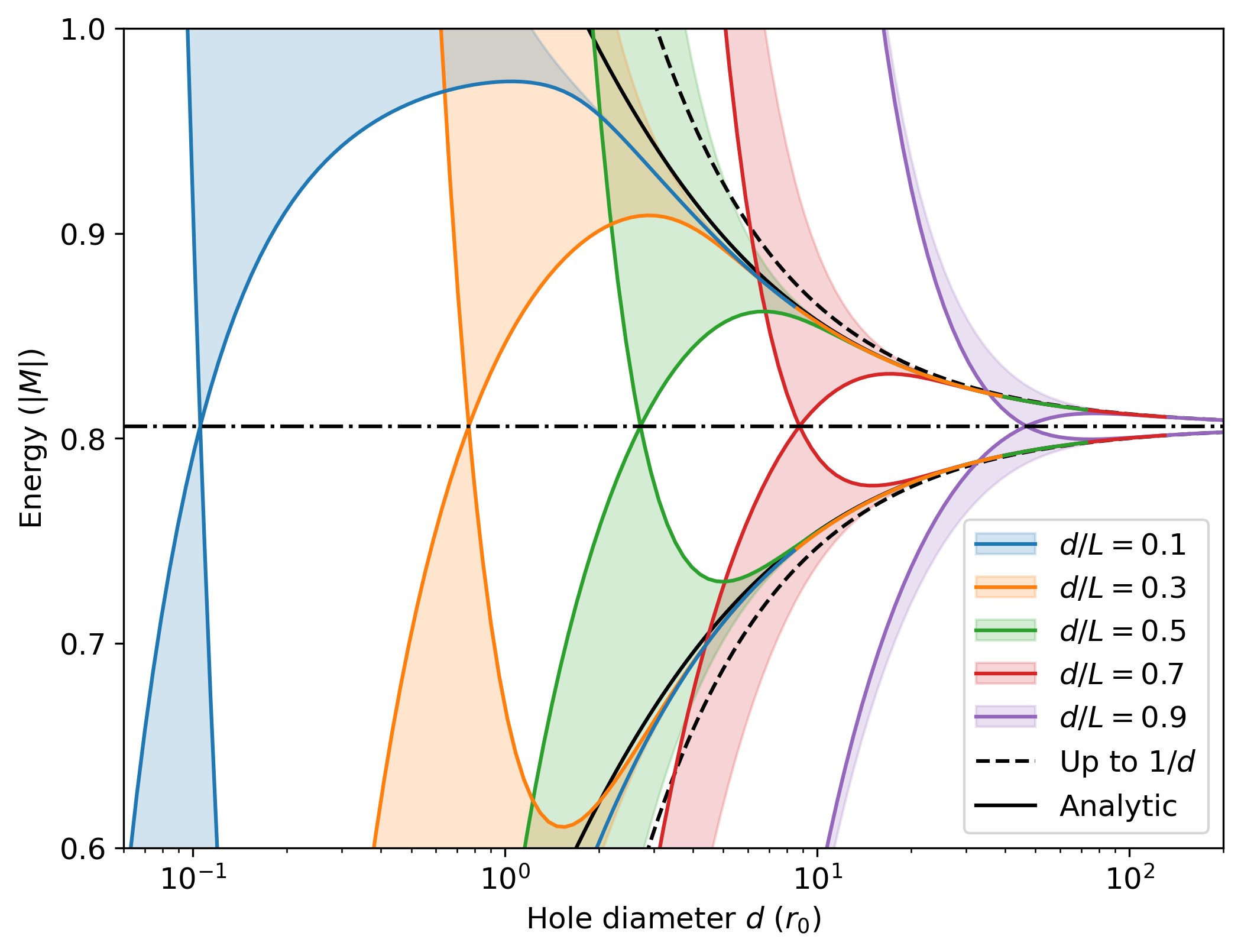}
    \caption{\label{fig:hex-analytic}
    The colored solid lines are the energies at the $\Gamma$ point of the two bands closest to Dirac point energy.
    The shaded region represents the difference in energy between the K and the $\Gamma$ points.
    The black solid lines are the analytic results given in Appendix~\ref{app:series} for $j = \pm 1/2$, and the black dashed lines are the same results truncated to the order of $1/d$ given by Eq.~\ref{eq:hole-energy}.
    The dash-dotted line is the energy of the Dirac point.}
\end{figure}

\subsection{\label{sec:results-localized}Localized states}

For systems as the ones considered in Fig.~\ref{fig:hex-state-b}, \ref{fig:hex-state-c}, and \ref{fig:hex-state-f}, the edge states around the holes are far apart and can be considered as isolated.
As these states are localized, they produce flat bands within the bulk gap in the band structure as shown by the solid straight lines in Fig.~\ref{fig:hex-bands-flat}.
Comparing to the band structure in the absence of holes (dotted lines), Fig.~\ref{fig:hex-bands-flat} shows that the introduction of holes perturbs the bulk band structure, as localized states (flat bands) emerge within the bulk gap.

Here we compare the energy of the localized states that we obtained numerically with the analytic result relative to an edge state around a single circular hole in a 2DTI.
The energy of this state in terms of the hole diameter $d$ is given by, see Appendix~\ref{app:analytic},
\begin{equation}
    E_{j,\uparrow/\downarrow} = E_0 \mp \frac{A\sqrt{B^2 - D^2}}{|B|} \frac{2j}{d} + O\left(\frac{1}{d^2}\right)
    \label{eq:hole-energy}
\end{equation}
up to first order in $1/d$,
where $A$, $B$, $C$, $D$, $M$ are the parameters defined in Section~\ref{sec:methods-bhz}, $E_0$ is the energy of the Dirac point in Eq.~\ref{eq:edge-energy}, and $j$ is a half-integer denoting the angular mode of the edge state.
In this study, the two states above and below the Dirac point correspond to $j = \pm 1/2$.
The series expansion of Eq.~\ref{eq:hole-energy} can be expressed to all orders, and regularized to give the exact result in terms of the exponential integral.
This exact result is shown as a solid line in Fig.~\ref{fig:hex-analytic}, and the detailed derivation is given in Appendix~\ref{app:series}.

In Fig.~\ref{fig:hex-analytic}, the energies of the band edges of the two bands closest to the Dirac point energy are shown as a function of $d$.
The bandwidth is shown by the shaded region, which shrinks in the limit of large $d$ (colored curves).
This shows the bands have no dispersion and become flat at that limit, which agrees with the localization of the states around holes.
The band energies also agree well with the analytic result at that limit.

The colored solid lines are observed to cross at the energy of the Dirac point for a specific hole radius that depends on the ratio $d/L$, indicating that the band gaps close at the Dirac point.
Each colored curve and shade represents a horizontal line on Fig.~\ref{fig:hex-phase}, where increasing $d$ in Fig.~\ref{fig:hex-analytic} corresponds to increasing $L$ in Fig.~\ref{fig:hex-phase}.
The system transitions from trivial to QSH phase as $d$ increases past the crossing point.
For example, focusing on the ratio $d/L = 0.1$ (blue in Fig.~\ref{fig:hex-analytic}), the crossing point (closing of the gap) occurs at $d \approx 0.1 r_0$, which corresponds to the transition point $L \approx r_0$ in Fig.~\ref{fig:hex-phase}.

Fig.~\ref{fig:hex-bands-confined} shows the band structure of the system corresponding to Fig.~\ref{fig:hex-state-d}, i.e.\ in the topologically trivial phase. Flat bands are still present, as in Fig.~\ref{fig:hex-bands-flat}, but in this case they correspond to bulk states that localize in the spaces between holes. When the holes are large, the space between the holes is pinched off, giving rise to triangular regions between holes.
The state is trapped in these regions as shown in Fig.~\ref{fig:hex-state-d}.
Note that, in Fig.~\ref{fig:hex-bands-confined}, the positions of the two closest flat bands to the Dirac point are not symmetric with respect to the Dirac point (the one above is actually beyond the bulk gap).

\begin{widetext}
\begin{figure*}[!h]
    \centering
    \includegraphics[width=\linewidth]{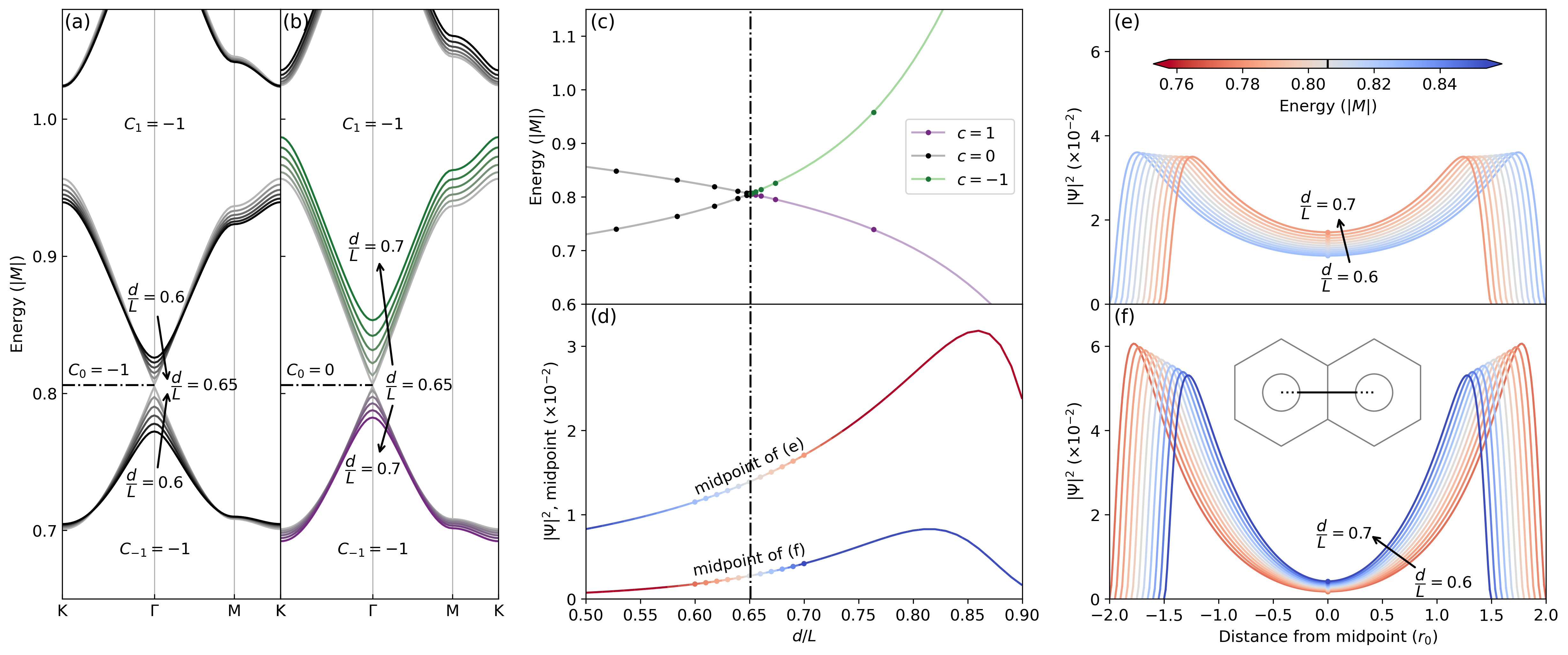}
    \phantomsubfloat{\label{fig:hex-bands-same-L-topological}}
    \phantomsubfloat{\label{fig:hex-bands-same-L-trivial}}
    \phantomsubfloat{\label{fig:hex-transition-same-L-energy}}
    \phantomsubfloat{\label{fig:hex-transition-same-L-prob}}
    \phantomsubfloat{\label{fig:hex-gap-same-L-bulk}}
    \phantomsubfloat{\label{fig:hex-gap-same-L-edge}}
    \caption{\label{fig:hex-transition-same-L}
    (a)(b) The probability density $|\Psi|^2$ of the two states above and below the Dirac point at the $\Gamma$ point.
    (a)(b) Overlaid band structures to show the transition in band structure (a) before the band gap closes and (b) after the band gap reopens, corresponding to the systems on the black solid line in Fig.~\ref{fig:hex-phase}.
    The bands of the same color intensity belong to the band structure of the same system.
    The color hue of the bands in (b) shows the Chern number of the band, which has the same color hue as in (c).
    The Chern number accumulated up to any gap $C$ is indicated in the gap.
    The dash-dotted line is the energy of the Dirac point.
    (c) The energy of the highest occupied and lowest unoccupied states.
    Color denotes the Chern number $c$ of the band.
    The solid dots are the sample points used during the golden section search, corresponding to those in Fig.~\ref{fig:hex-phase}.
    (d) The probability density $|\Psi|^2$ at the midpoint between adjacent holes (i.e.\ the zero point on the horizontal axis of (e) and (f)).
    The solid dots correspond to the values at the midpoints indicated by solid dots in (e) and (f).
    The dash-dotted line across (c) and (d) is the transition point.
    (e)(f) The probability density $|\Psi|^2$ of the two states above and below the Dirac point at the $\Gamma$ point.
    $|\Psi|^2$ is normalized for the total probability within a supercell to be 1.
    (e) and (f) are plotted along the line connecting the center of adjacent holes (the black solid line in the inset) for different $d/L$ ratios, where $L = 10 r_0$.
    The solid mark on the energy color bar in (e) is the energy of the Dirac point, which corresponds to the dash-dotted line across (a) and (b).}
\end{figure*}
\end{widetext}

\subsection{\label{sec:results-transition}Transition between the QSH and trivial phases}

The transition along the black solid line at $L = 10 r_0$ in Fig.~\ref{fig:hex-phase} is investigated extensively in Fig.~\ref{fig:hex-transition-same-L}.
The band structures for $d/L \leq 0.65$ and $d/L \geq 0.65$ are presented in Fig.~\ref{fig:hex-bands-same-L-topological} and \ref{fig:hex-bands-same-L-trivial}, which show that the band gap closes at $d/L = 0.65$.
By calculating the Chern number of each of the two bands, we find that they gain opposite Chern numbers when the band gap reopens, yielding zero for the Chern number accumulated up to the band gap.

The $\Gamma$ point energies of the top valence band and the bottom conduction band are plotted in Fig.~\ref{fig:hex-transition-same-L-energy}, which shows two curves crossing at $d/L = 0.65$.
This hints that the two states with energies above and below the Dirac point at the $\Gamma$ point are exchanged, at the closing of the band gap, between the valence and conduction bands, as confirmed by the plots of their probability densities in Fig.~\ref{fig:hex-gap-same-L-bulk} and \ref{fig:hex-gap-same-L-edge}.

This shows that the exchange of states between the two bands changes their topology, and demonstrates the mechanism of the topological phase transition induced by increasing the hole size.

The probability densities at the midpoint in the space between holes for the two states are plotted in Fig.~\ref{fig:hex-transition-same-L-prob}.
The midpoint probability density is significantly lower for the state in Fig.~\ref{fig:hex-gap-same-L-edge} than for that in Fig.~\ref{fig:hex-gap-same-L-bulk}.
The low probability density at the midpoint in comparison to the peaks on the side in Fig.~\ref{fig:hex-gap-same-L-edge} shows that this state is localized around the holes.
In contrast, the probability density at the midpoint is comparable to the peaks in Fig.~\ref{fig:hex-gap-same-L-bulk}, showing that this state is delocalized in the spaces between the holes.

Finally, the fact that the midpoint probability density decreases for both states when $d/L > 0.86$ in Fig.~\ref{fig:hex-transition-same-L-prob} also shows that the pinching off expels the state from the space between the holes.

\section{\label{sec:conclusion}Conclusion}

In this work, we used the finite element method to model 2D topological insulators, described by the BHZ model, patterned with a hexagonal superlattice of circular holes.
We calculated the topological phase diagram to explore the interplay between the hole period and diameter.
We found that altering the hole size and spacing modifies the topological property of the material.
When the holes are well-separated, they can be treated as isolated and the states on either side of the band gap are localized around the holes, creating flat bands in the band structure.
When the separation between the holes is smaller than a certain fraction of the hole period, the edge states overlap significantly and the material ceases to be a topological insulator.
The feature sizes needed to achieve a phase transition are on the order of hundreds of nanometers, which can be fabricated with current lithographic techniques.
This prompts future experiments to test the result of this study.

This work also demonstrates the feasibility of using the finite element method in modeling topological insulators.
A symmetric geometry was studied in this work, but the approach is directly applicable to irregular geometries.
This includes modeling 3D topological insulators with superlattice structures.
In the future, the model may be extended to study electronic transport in topological insulators, enabling the study of their behavior in the context of electronic devices.

\appendix%*

\section{\label{app:analytic}Analytic solution for the edge states}

With inspiration from the literature \cite{Wada2011,Gioia2019,Governale2020,Governale2023}, we derive the helical edge states around a hole of radius $R$ and their energies.

Let us work in polar coordinates, so the state exists in the region $r > R$.
Then, with
\begin{align}
    \partial_x &= \cos\phi \, \partial_r - \frac{\sin\phi}{r} \partial_\phi, \\
    \partial_y &= \sin\phi \, \partial_r + \frac{\cos\phi}{r} \partial_\phi,
\end{align}
the Laplacian is
\begin{equation}
    \partial^2 = \partial_x^2 + \partial_y^2 = \partial_r^2 + \frac{1}{r} \partial_r + \frac{1}{r^2} \partial_\phi^2 .
\end{equation}
For convenience, we also calculate
\begin{align}
    \partial_\pm &= \partial_x \pm i\partial_y \nonumber \\
    &= (\cos\phi \pm i\sin\phi) \partial_r + \frac{1}{r} (-\sin\phi \pm i\cos\phi) \partial_\phi \nonumber \\
    &= e^{\pm i\phi} \partial_r + \frac{1}{r} e^{\pm i\left(\phi+\frac{\pi}{2}\right)} \partial_\phi \nonumber \\
    &= e^{\pm i\phi} \left(\partial_r \pm \frac{i}{r} \partial_\phi\right) .
\end{align}
Hence, the wavenumber operators can be expressed as
\begin{align}
    k^2 &= -\partial^2 = -\left(\partial_r^2 + \frac{1}{r} \partial_r + \frac{1}{r^2} \partial_\phi^2\right) , \\
    k_\pm &= -i\partial_\pm = e^{\pm i\phi} \left(-i\partial_r \pm \frac{1}{r} \partial_\phi\right) .
\end{align}

Consider the upper left block of the BHZ Hamiltonian, which can be written as
\begin{align}
    H_\uparrow &= \begin{pmatrix}
        M_+ - B_+ k^2 & A k_+ \\
        A k_- & -M_- + B_- k^2
    \end{pmatrix} \nonumber \\
    &= \begin{pmatrix}
        M_+ + B_+ \partial^2 & A e^{i\phi} \partial_+ \\
        A e^{-i\phi} \partial_- & -M_- - B_- \partial^2
    \end{pmatrix} ,
\end{align}
with $M_\pm = M \pm C$ and $B_\pm = B \pm D$.
Let us solve the Schr\"{o}dinger equation $H_\uparrow \Psi = E_\uparrow \Psi$.
By using the ansatz
\begin{equation}
    \Psi(r,\phi) = \frac{e^{ij\phi}}{\sqrt{2\pi r}} \begin{pmatrix} e^{i\phi/2} u_1(r) \\ e^{-i\phi/2} u_2(r) \end{pmatrix} ,
\end{equation}
with \(j\) half-integer to respect the \(2\pi\) periodicity of \(\phi\), we find
\begin{equation}
    H_{\text{eff},j} \Phi = E_\uparrow \Phi
\end{equation}
where the Hamiltonian can be split as $H_{\text{eff},j} = H_0 + H_{1,j}$, with
\begin{align}
    H_0 &= \begin{pmatrix}
        M_+ + B_+ \partial_r^2 & -iA \partial_r \\
        -iA \partial_r & -M_- - B_- \partial_r^2
    \end{pmatrix} , \\
    H_{1,j} &= \begin{pmatrix}
        -B_+ \frac{j(j+1)}{r^2} & iA \frac{j}{r} \\
        -iA \frac{j}{r} & B_- \frac{j(j-1)}{r^2}
    \end{pmatrix} ,
\end{align}
and $\Phi$ takes the form of
\begin{equation}
    \Phi = \begin{pmatrix} u_1(r) \\ u_2(r) \end{pmatrix}
\end{equation}
and is normalized as
\begin{equation}
    \int_R^\infty \mathrm{d}r\, \Phi^\dagger(r) \cdot \Phi(r) = 1 .
\end{equation}

Let us start calculating the eigenstates $\Phi_0$ and eigenvalues $E_0$ of $H_0$, which corresponds to the limit $R \to \infty$:
\begin{equation}
    H_0 \Phi_0 = E_0 \Phi_0.
\end{equation}
By substituting the ansatz
\begin{equation}
    \Phi_0(r) = \begin{pmatrix} u \\ v \end{pmatrix} e^{\lambda (r-R)} ,
\end{equation}
we obtain
\begin{equation}
    \begin{pmatrix}
        M_+ + B_+ \lambda^2 - E_0 & -iA \lambda \\
        -iA \lambda & -M_- - B_- \lambda^2 - E_0
    \end{pmatrix} \begin{pmatrix} u \\ v \end{pmatrix} = 0 .
\end{equation}
Non-trivial solutions exist when the determinant of the matrix is zero, i.e.\ for $\lambda = \lambda_{1,2}$ where
\begin{equation}
    \lambda_{1,2} = \sqrt{\frac{\tilde{A} \pm \sqrt{\tilde{A}^2 + 4\tilde{M}\tilde{B}}}{2\tilde{B}}}
    \label{eq:lambda}
\end{equation}
where
\begin{align}
    \tilde{A} &= A^2 - M_+ B_- - M_- B_+ - E_0 (B_+ - B_-) \nonumber \\
    &= A^2 - 2MB + 2CD - 2DE_0 , \\
    \tilde{M} &= (E_0 - M_+)(E_0 + M_-) \nonumber \\
    &= (E_0 - C)^2 - M^2 , \\
    \tilde{B} &= B_+ B_- = B^2 - D^2 .
\end{align}
The eigenstates relative to $\pm\lambda_1$ are
\begin{equation}
    u_{1,\pm} = \frac{1}{L_1} \begin{pmatrix} \pm \frac{iA\lambda_1}{M_+ + B_+ \lambda_1^2 - E_0} \\ 1 \end{pmatrix}
\end{equation}
and those for $\pm\lambda_2$ are
\begin{equation}
    u_{2,\pm} = \frac{1}{L_2} \begin{pmatrix} \pm \frac{iA\lambda_2}{M_+ + B_+ \lambda_2^2 - E_0} \\ 1 \end{pmatrix} ,
\end{equation}
where $L_1$ and $L_2$ are normalization factors such that $|u_{i,\pm}|^2 = 1$.
The general wavefunction for the Schr\"{o}dinger equation for $H_0$ can be written as the superposition
\begin{align}
    \Phi_0(r) &= a e^{\lambda_1(r-R)} u_{1,+} + b e^{\lambda_2(r-R)} u_{2,+} \nonumber \\
    &+ c e^{-\lambda_1(r-R)} u_{1,-} + d e^{-\lambda_2(r-R)} u_{2,-} .
\end{align}
The boundary condition at infinity requires $\Phi_0(r) \to 0$ for $r \to \infty$.
Assuming $\operatorname{Re}(\lambda_1) > 0$ and $\operatorname{Re}(\lambda_2) > 0$, then we have to set $a = b = 0$.
Imposing the boundary condition at the border of the hole [$\Phi_0(r=R) = 0$] gives
\begin{equation}
    c u_{1,-} + d u_{2,-} = 0 ,
\end{equation}
which yields
\begin{equation}
    \frac{\lambda_1}{M_+ + B_+ \lambda_1^2 - E_0} = \frac{\lambda_2}{M_+ + B_+ \lambda_2^2 - E_0}.
\end{equation}
By substituting the expressions for $\lambda_1$ and $\lambda_2$ in \ref{eq:lambda}, we find
\begin{equation}
    E_0 = \frac{M_+ B_- - M_- B_+}{B_+ + B_-} = C - \frac{MD}{B}
\end{equation}
and
\begin{equation}
    \Phi_0(r) = \rho(r) u_-
\end{equation}
where
\begin{equation}
    u_- \equiv u_{1,-} = u_{2,-} = \frac{1}{\sqrt{2}} \begin{pmatrix} -i\sqrt{B_-/|B|} \\ \sqrt{B_+/|B|} \end{pmatrix} 
\end{equation}
and
\begin{equation}
    \rho(r) = \frac{1}{N} \left[e^{-\lambda_1(r-R)} - e^{-\lambda_2(r-R)}\right]
\end{equation}
with $N$ ensuring normalization:
\begin{align}
    N^2 &= \int_R^\infty \mathrm{d}r\, \left|e^{-\lambda_1(r-R)} - e^{-\lambda_2(r-R)}\right|^2 \nonumber \\
    &= \int_0^\infty \mathrm{d}r\, \left(e^{-2\lambda_1 r} + e^{-2\lambda_2 r} - 2e^{-(\lambda_1+\lambda_2)r}\right) \nonumber \\
    &= \frac{1}{2\lambda_1} + \frac{1}{2\lambda_2} - \frac{2}{\lambda_1+\lambda_2} \nonumber \\
    &= \frac{(\lambda_1 - \lambda_2)^2}{2\lambda_1 \lambda_2 (\lambda_1 + \lambda_2)}
    \label{eq:norm}
\end{align}
so that
\begin{equation}
    \int_R^\infty \mathrm{d}r\, |\rho(r)|^2 = 1 .
\end{equation}

For convenience, we simplify the following quantities with $E_0 = C - MD/B$:
\begin{align}
    \tilde{A} &= A^2 - 2MB + 2CD - 2D\left(C - \frac{MD}{B}\right) \nonumber \\
    &= A^2 - 2MB + \frac{2MD^2}{B} \nonumber \\
    &= A^2 - 2M\frac{B_+ B_-}{B} , \\
    \tilde{M} &= \left(C - \frac{MD}{B} - C\right)^2 - M^2 \nonumber \\
    &= -M^2 \frac{B_+ B_-}{B^2} , \\
    \lambda_1 \lambda_2 &= \sqrt{\frac{\tilde{A}^2 - (\tilde{A}^2 + 4\tilde{M}\tilde{B})}{4\tilde{B}^2}} = \sqrt{-\frac{\tilde{M}}{\tilde{B}}} = \frac{M}{B} , \label{eq:lambda-prod} \\
    (\lambda_1 \pm \lambda_2)^2 &= \lambda_1^2 + \lambda_2^2 \pm 2\lambda_1 \lambda_2 = \frac{\tilde{A}}{\tilde{B}} \pm \frac{2M}{B} \nonumber \\
    &= \frac{A^2}{B_+ B_-} - \frac{2M}{B} \pm \frac{2M}{B} , \label{eq:lambda-sum} \\
    \lambda_{1,2} &= \frac{(\lambda_1 + \lambda_2) \pm (\lambda_1 - \lambda_2)}{2} \nonumber \\
    &= \frac{A}{2\sqrt{B_+ B_-}} \pm \sqrt{\frac{A^2}{4B_+ B_-} - \frac{M}{B}} .
\end{align}
For the material to be a TI \cite{BHZ} and for real solutions of $\lambda$ to exist \cite{Zhou2008}, we require
\begin{equation}
    \frac{A^2}{4B_+ B_-} = \frac{A^2}{4(B^2 - D^2)} > \frac{M}{B} > 0 .
\end{equation}
Without loss of generality, we take $A > 0$.
The normalization coefficient can then be simplified as
\begin{align}
    N^2 &= \frac{(\lambda_1 - \lambda_2)^2}{2\lambda_1 \lambda_2 (\lambda_1 + \lambda_2)} \nonumber \\
    &= \left(\frac{A^2}{B_+ B_-} - \frac{4M}{B}\right) / \left(\frac{2M}{B} \frac{A}{\sqrt{B_+ B_-}}\right) \nonumber \\
    &= \frac{B}{2M} \frac{A}{\sqrt{B_+ B_-}} - \frac{2\sqrt{B_+ B_-}}{A} .
\end{align}

$\lambda_2 < \lambda_1$, so $\lambda_2$ sets the decay length of $\rho(r)$.
The series expansion of the square root in $\lambda_2$ gives
\begin{align}
    \lambda_2 &= \frac{A}{2\sqrt{B_+ B_-}} \nonumber \\
    &- \left(\frac{A}{2\sqrt{B_+ B_-}} - \frac{2\sqrt{B_+ B_-}}{A} \frac{M}{2B}\right) + O\left(\frac{M^2}{B^2}\right) \nonumber \\
    &= \frac{M}{A} \frac{\sqrt{B_+ B_-}}{B} + O\left(\frac{M^2}{B^2}\right) .
\end{align}
Therefore, the decay length is
\begin{equation}
    \lambda_2^{-1} \approx \frac{A}{M} \frac{B}{\sqrt{B_+ B_-}}.
\end{equation}
$B/\sqrt{B_+ B_-}$ contains the dependence of $D$, and reduces to $\operatorname{sgn}(B)$ when $D=0$.
This prompts us to use $r_0 = A/|M|$ as the length scale.

Let us now calculate the correction to the energy $E_0$ coming from the contribution of the Hamiltonian $H_{1,j}$.
Using perturbation theory, we calculate the matrix element
\begin{align}
    E_j &= \braket{\Phi_0|H_{1,j}|\Phi_0} \nonumber \\
    &= \int_R^\infty \mathrm{d}r\, \frac{|\rho(r)|^2}{|B|} \left(-A\sqrt{B_+ B_-}\frac{j}{r} - B_+ B_-\frac{j}{r^2}\right) \nonumber \\
    &= -\frac{A\sqrt{B_+ B_-}}{|B|} j F(R) - \frac{B_+ B_-}{|B|} j G(R)
    \label{eq:Ej}
\end{align}
where, expanding up to first order in $1/R$ and for sufficiently large $R$,
\begin{align}
    F(R) &= \int_R^\infty \mathrm{d}r\, \frac{|\rho(r)|^2}{r} \nonumber \\
    &= \int_0^\infty \mathrm{d}r\, \frac{|\rho(r+R)|^2}{r+R} \nonumber \\
    &= \int_0^\infty \mathrm{d}r\, |\rho(r+R)|^2 \left(\frac{1}{R} + O\left(\frac{1}{R^2}\right)\right) \nonumber \\
    &= \frac{1}{R} + O\left(\frac{1}{R^2}\right) , \label{eq:appendix-F} \\
    G(R) &= \int_R^\infty \mathrm{d}r\, \frac{|\rho(r)|^2}{r^2} \nonumber \\
    &= \int_0^\infty \mathrm{d}r\, \frac{|\rho(r+R)|^2}{(r+R)^2} \nonumber \\
    &= \int_0^\infty \mathrm{d}r\, |\rho(r+R)|^2 \left(\frac{1}{R^2} + O\left(\frac{1}{R^3}\right)\right) \nonumber \\
    &= 0 + O\left(\frac{1}{R^2}\right) . \label{eq:appendix-G}
\end{align}
Therefore,
\begin{align}
    E_j &= -\frac{A\sqrt{B_+ B_-}}{|B|} \frac{j}{R} + O\left(\frac{1}{R^2}\right) \nonumber \\
    &= -\frac{A\sqrt{B^2 - D^2}}{|B|} \frac{j}{R} + O\left(\frac{1}{R^2}\right)
\end{align}
and
\begin{align}
    E_\uparrow &= E_0 + E_j \nonumber \\
    &= C - \frac{MD}{B} - \frac{A\sqrt{B^2 - D^2}}{|B|} \frac{j}{R} + O\left(\frac{1}{R^2}\right) .
\end{align}

Regarding the lower-right block of the BHZ Hamiltonian being the time-reversal of the upper-left block, we expect to find
\begin{align}
    E_\downarrow &= E_0 - E_j \nonumber \\
    &= C - \frac{MD}{B} + \frac{A\sqrt{B^2 - D^2}}{|B|} \frac{j}{R} + O\left(\frac{1}{R^2}\right) .
\end{align}
Putting these two expressions in terms of $d$ gives the expression in Section~\ref{sec:results-localized}.

\begin{widetext}
    
    \section{\label{app:series}Full series expansion of the analytic solution}

Expanding $1/(r+R)$ and $1/(r+R)^2$ about $R\to\infty$ gives
\begin{align}
    \frac{1}{r+R} &= \sum_{n=1}^\infty \frac{(-r)^{n-1}}{R^n} , \\
    \frac{1}{(r+R)^2} &= \sum_{n=1}^\infty \frac{(n-1)(-r)^{n-2}}{R^n} .
\end{align}
The expansion in Eq.~\ref{eq:appendix-F} and \ref{eq:appendix-G} can be written out fully as
\begin{align}
    F(R) &= \int_0^\infty \mathrm{d}r\, \frac{|\rho(r+R)|^2}{r+R} = \sum_{n=1}^\infty \frac{(-1)^{n-1} I_{n-1}}{R^n} , \\
    G(R) &= \int_0^\infty \mathrm{d}r\, \frac{|\rho(r+R)|^2}{(r+R)^2} = \sum_{n=2}^\infty \frac{(n-1)(-1)^{n-2} I_{n-2}}{R^n} ,
\end{align}
where
\begin{align}
    I_n &= \int_0^\infty \mathrm{d}r\, r^n |\rho(r+R)|^2 \nonumber \\
    &= \frac{1}{N^2} \int_0^\infty \mathrm{d}r\, r^n \left|e^{-\lambda_1 r} - e^{-\lambda_2 r}\right|^2 \nonumber \\
    &= \frac{1}{N^2} \int_0^\infty \mathrm{d}r\, r^n \left(e^{-2\lambda_1 r} + e^{-2\lambda_2 r} - 2e^{-(\lambda_1 + \lambda_2)r}\right) \nonumber \\
    &= \frac{1}{N^2} \left(\frac{n!}{(2\lambda_1)^{n+1}} + \frac{n!}{(2\lambda_2)^{n+1}} - 2\frac{n!}{(\lambda_1 + \lambda_2)^{n+1}}\right) \nonumber \\
    &= \frac{2\lambda_1 \lambda_2 (\lambda_1 + \lambda_2)}{(\lambda_1 - \lambda_2)^2} n!\left(\frac{\lambda_1^{n+1} + \lambda_2^{n+1}}{(2\lambda_1 \lambda_2)^{n+1}} - \frac{2}{(\lambda_1 + \lambda_2)^{n+1}}\right) \nonumber \\
    &= \frac{n!}{(\lambda_1 - \lambda_2)^2} \left(\frac{(\lambda_1 + \lambda_2)(\lambda_1^{n+1} + \lambda_2^{n+1})}{(2\lambda_1 \lambda_2)^n} - \frac{4\lambda_1 \lambda_2}{(\lambda_1 + \lambda_2)^n}\right) \nonumber \\
    &= \frac{n!}{(\lambda_1 - \lambda_2)^2} (a_n - b_n)
\end{align}
with
\begin{align}
    a_n &= \frac{(\lambda_1 + \lambda_2)(\lambda_1^{n+1} + \lambda_2^{n+1})}{(2\lambda_1 \lambda_2)^n} , \\
    b_n &= \frac{4\lambda_1 \lambda_2}{(\lambda_1 + \lambda_2)^n} .
\end{align}
Eq.~\ref{eq:Ej} can be factored as
\begin{align}
    E_j &= -\frac{A\sqrt{B_+ B_-}}{|B|} j \left(F(R) + \frac{\sqrt{B_+ B_-}}{A} G(R)\right) \nonumber \\
    &= -\frac{A\sqrt{B_+ B_-}}{|B|} j \left(F(R) + \frac{G(R)}{\lambda_1 + \lambda_2}\right) .
\end{align}
If we write
\begin{align}
    E_j &= -\frac{A\sqrt{B_+ B_-}}{|B|} j \sum_{n=1}^\infty \frac{S_n}{R^n} , \\
    F(R) &= \sum_{n=1}^\infty \frac{F_n}{R^n} , \\
    G(R) &= \sum_{n=1}^\infty \frac{G_n}{R^n} ,
\end{align}
the contribution to $E_j$ of each order of $1/R$ can be calculated by
\begin{align}
    S_n &= F_n + \frac{G_n}{\lambda_1 + \lambda_2} \nonumber \\
    &= (-1)^{n-1} I_{n-1} + \frac{1}{\lambda_1 + \lambda_2} (n-1)(-1)^{n-2} I_{n-2} \nonumber \\
    &= \frac{(-1)^{n-1} (n-1)!}{(\lambda_1 - \lambda_2)^2} \left((a_{n-1} - b_{n-1}) - \frac{1}{\lambda_1 + \lambda_2} (a_{n-2} - b_{n-2})\right) \nonumber \\
    &= \frac{(-1)^{n-1} (n-1)!}{(\lambda_1 - \lambda_2)^2} \left(\left(a_{n-1} - \frac{a_{n-2}}{\lambda_1 + \lambda_2}\right) - \left(b_{n-1} - \frac{b_{n-2}}{\lambda_1 + \lambda_2}\right)\right) ,
\end{align}
where
\begin{align}
    a_{n-1} - \frac{a_{n-2}}{\lambda_1 + \lambda_2} &= (2\lambda_1 \lambda_2)^{-n+1} \left((\lambda_1 + \lambda_2)(\lambda_1^n + \lambda_2^n) - 2\lambda_1\lambda_2(\lambda_1^{n-1} + \lambda_2^{n-1})\right) \nonumber \\
    &= (2\lambda_1 \lambda_2)^{-n+1} \left(\lambda_1^{n+1} + \lambda_2^{n+1} + \lambda_1^n \lambda_2 + \lambda_1 \lambda_2^n - 2\lambda_1^n \lambda_2 - 2\lambda_1 \lambda_2^n\right) \nonumber \\
    &= (2\lambda_1 \lambda_2)^{-n+1} \left(\lambda_1^{n+1} + \lambda_2^{n+1} - \lambda_1^n \lambda_2 - \lambda_1 \lambda_2^n\right) \nonumber \\
    &= (2\lambda_1 \lambda_2)^{-n+1} (\lambda_1 - \lambda_2)(\lambda_1^n - \lambda_2^n) \nonumber \\
    &= (2\lambda_1 \lambda_2)^{-n+1} (\lambda_1 - \lambda_2)^2 \sum_{i=0}^{n-1} \lambda_1^i \lambda_2^{n-1-i} , \\
    b_{n-1} - \frac{b_{n-2}}{\lambda_1 + \lambda_2} &= \frac{4\lambda_1 \lambda_2}{(\lambda_1 + \lambda_2)^{n-1}} - \frac{4\lambda_1 \lambda_2}{(\lambda_1 + \lambda_2)^{n-2+1}} = 0 .
\end{align}
Therefore,
\begin{equation}
    S_n = \frac{(-1)^{n-1} (n-1)!}{(2\lambda_1 \lambda_2)^{n-1}} \sum_{i=0}^{n-1} \lambda_1^i \lambda_2^{n-1-i} .
    \label{eq:coefficient}
\end{equation}
Taking $n = 1$ immediately gives $S_1 = 1$, which is consistent with the result before.
Calculating the next few terms:
\begin{align}
    S_2 &= \frac{-(1!)}{2\lambda_1 \lambda_2} (\lambda_1 + \lambda_2) = -\frac{B}{2M} \frac{A}{\sqrt{B_+ B_-}} , \\
    S_3 &= \frac{(2!)}{(2\lambda_1 \lambda_2)^2} (\lambda_1^2 + \lambda_1 \lambda_2 + \lambda_2^2) \nonumber \\
    &= \frac{2}{(2\lambda_1 \lambda_2)^2} \left((\lambda_1 + \lambda_2)^2 - \lambda_1 \lambda_2\right) \nonumber \\
    &= 2\left(\frac{B}{2M}\right)^2 \left(\frac{A^2}{B_+ B_-} - \frac{M}{B}\right) , \\
    S_4 &= \frac{-(3!)}{(2\lambda_1 \lambda_2)^3} (\lambda_1^3 + \lambda_1^2 \lambda_2 + \lambda_1 \lambda_2^2 + \lambda_2^3) \nonumber \\
    &= \frac{-6}{(2\lambda_1 \lambda_2)^3} \left((\lambda_1 + \lambda_2)^3 - 2\lambda_1 \lambda_2 (\lambda_1 + \lambda_2)\right) \nonumber \\
    &= \frac{-6}{(2\lambda_1 \lambda_2)^3} (\lambda_1 + \lambda_2) \left((\lambda_1 + \lambda_2)^2 - 2\lambda_1 \lambda_2\right) \nonumber \\
    &= -6 \left(\frac{B}{2M}\right)^3 \frac{A}{\sqrt{B_+ B_-}} \left(\frac{A^2}{B_+ B_-} - \frac{2M}{B}\right) .
\end{align}

\begin{figure*}[!t]
    \centering
    \includegraphics[width=0.8\linewidth]{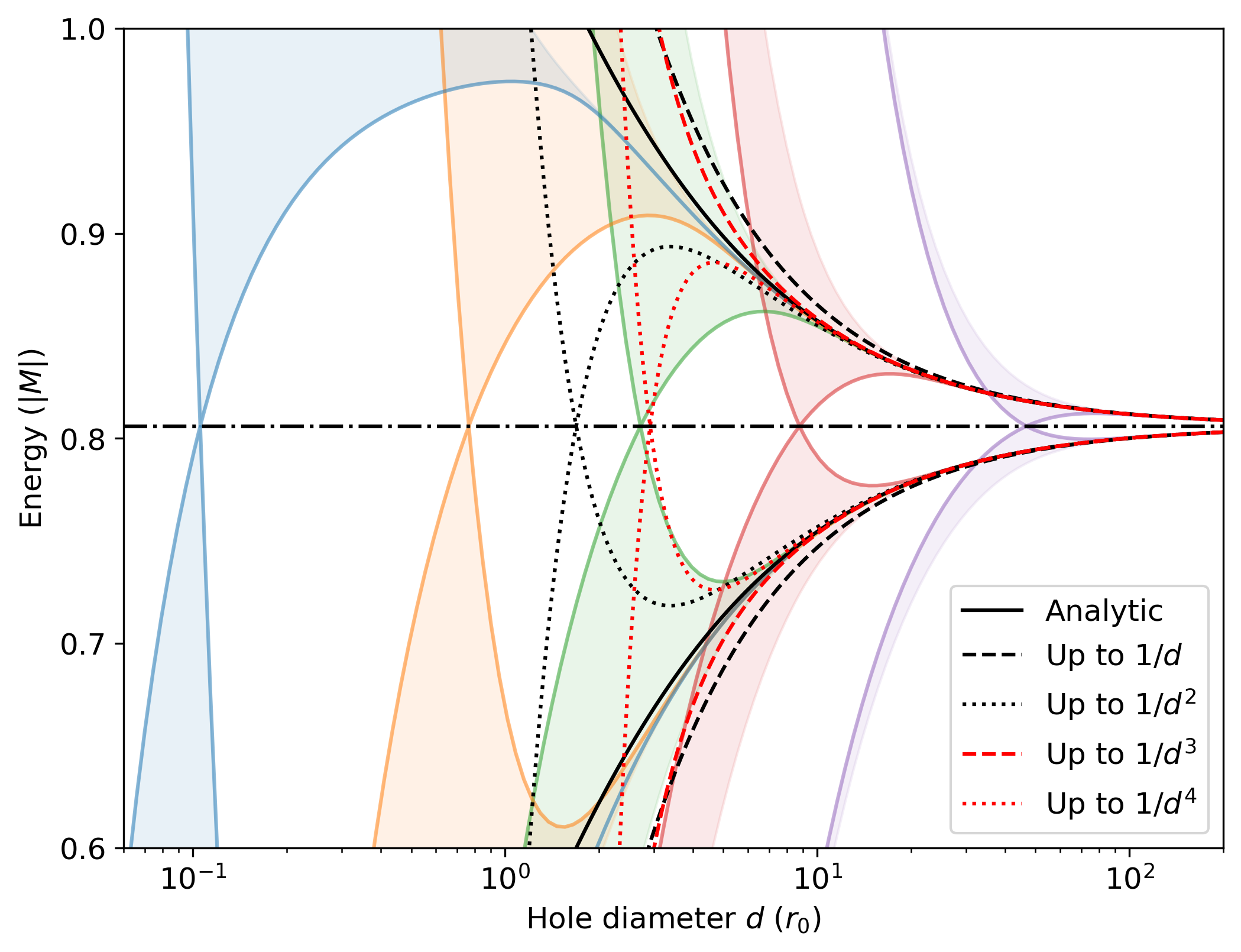}
    \caption{\label{fig:hex-analytic-series}
    The energy of the edge states around holes versus hole radii.
    The colored solid lines are the energies at the $\Gamma$ point.
    The shaded region represents the difference in energy between the K and the $\Gamma$ points.
    The colors are the same as in Fig.~\ref{fig:hex-analytic}.
    The black solid lines are the analytic results given by Eq.~\ref{eq:analytic} for $j = \pm 1/2$, and the dashed lines are the same results truncated to different orders of $1/R$ computed with Eq.~\ref{eq:coefficient}.
    The dash-dotted line is the energy of the Dirac point.}
\end{figure*}

Taking the first four terms of the expansion gives
\begin{align}
    E_j = -\frac{A\sqrt{B_+ B_-}}{|B|} j \Biggl(&\frac{1}{R} - \frac{B}{2M} \frac{A}{\sqrt{B_+ B_-}} \frac{1}{R^2} + \left(\frac{B}{2M}\right)^2 \left(\frac{A^2}{B_+ B_-} - \frac{M}{B}\right) \frac{2}{R^3} \nonumber \\
    &- \left(\frac{B}{2M}\right)^3 \frac{A}{\sqrt{B_+ B_-}} \left(\frac{A^2}{B_+ B_-} - \frac{2M}{B}\right) \frac{6}{R^4} \Biggr) + O\left(\frac{1}{R^5}\right) .
\end{align}
Written in terms of $d$ gives
\begin{align}
    E_j = -\frac{A\sqrt{B_+ B_-}}{|B|} 2j \Biggl(&\frac{1}{d} - \frac{B}{M} \frac{A}{\sqrt{B_+ B_-}} \frac{1}{d^2} + \left(\frac{B}{M}\right)^2 \left(\frac{A^2}{B_+ B_-} - \frac{M}{B}\right) \frac{2}{d^3} \nonumber \\
    &- \left(\frac{B}{M}\right)^3 \frac{A}{\sqrt{B_+ B_-}} \left(\frac{A^2}{B_+ B_-} - \frac{2M}{B}\right) \frac{6}{d^4} \Biggr) + O\left(\frac{1}{d^5}\right) .
\end{align}

$S_n$ can also be calculated to be
\begin{align}
    S_n &= \frac{(-1)^{n-1} (n-1)!}{(\lambda_1 - \lambda_2)^2 (2\lambda_1 \lambda_2)^{n-1}} \left((\lambda_1 + \lambda_2)(\lambda_1^n + \lambda_2^n) - 2\lambda_1\lambda_2(\lambda_1^{n-1} + \lambda_2^{n-1})\right) \nonumber \\
    &= \frac{(-1)^{n-1} (n-1)!}{(\lambda_1 - \lambda_2)^2 (2\lambda_1 \lambda_2)^{n-1}} \left(((\lambda_1 + \lambda_2)\lambda_1 - 2\lambda_1\lambda_2) \lambda_1^{n-1} + ((\lambda_1 + \lambda_2)\lambda_2 - 2\lambda_1\lambda_2) \lambda_2^{n-1}\right) \nonumber \\
    &= \frac{(-1)^{n-1} (n-1)!}{(\lambda_1 - \lambda_2)^2} \left((\lambda_1^2 - \lambda_1\lambda_2) \left(\frac{\lambda_1}{2\lambda_1 \lambda_2}\right)^{n-1} + (\lambda_2^2 - \lambda_1\lambda_2) \left(\frac{\lambda_2}{2\lambda_1 \lambda_2}\right)^{n-1}\right) \nonumber \\
    &= \frac{(-1)^{n-1} (n-1)!}{(\lambda_1 - \lambda_2)^2} \left(\lambda_1 (\lambda_1 - \lambda_2) \left(\frac{1}{2\lambda_2}\right)^{n-1} - \lambda_2 (\lambda_1 - \lambda_2) \left(\frac{1}{2\lambda_1}\right)^{n-1}\right) \nonumber \\
    &= \frac{(-1)^{n-1} (n-1)!}{\lambda_1 - \lambda_2} \left(\lambda_1 \left(\frac{1}{2\lambda_2}\right)^{n-1} - \lambda_2 \left(\frac{1}{2\lambda_1}\right)^{n-1}\right) .
\end{align}
We recognize the divergent series of
\begin{equation}
    x e^x E_1(x) = x e^x \int_x^\infty \frac{e^{-t}}{t} \,\mathrm{d}t = \sum_{n=0}^\infty \frac{n!}{(-x)^n}
\end{equation}
where $E_1(x) = \int_x^\infty e^{-t}/t \,\mathrm{d}t$ is the exponential integral, which is regularized with Borel summation.
The analytic solution can then be written as
\begin{align}
    E_j &= -\frac{A\sqrt{B_+ B_-}}{|B|} j \sum_{n=1}^\infty \frac{(-1)^{n-1} (n-1)!}{(\lambda_1 - \lambda_2) R^n} \left(\lambda_1 \left(\frac{1}{2\lambda_2}\right)^{n-1} - \lambda_2 \left(\frac{1}{2\lambda_1}\right)^{n-1}\right) \nonumber \\
    &= -\frac{A\sqrt{B_+ B_-}}{|B|} j \sum_{n=0}^\infty \frac{(-1)^n n!}{(\lambda_1 - \lambda_2) R^{n+1}} \left(\lambda_1 \left(\frac{1}{2\lambda_2}\right)^n - \lambda_2 \left(\frac{1}{2\lambda_1}\right)^n\right) \nonumber \\
    &= -\frac{A\sqrt{B_+ B_-}}{|B|} \frac{j}{(\lambda_1 - \lambda_2)R} \sum_{n=0}^\infty n! \left(\lambda_1 \left(\frac{1}{-2\lambda_2 R}\right)^n - \lambda_2 \left(\frac{1}{-2\lambda_1 R}\right)^n\right) \nonumber \\
    &= -\frac{A\sqrt{B_+ B_-}}{|B|} \frac{j}{(\lambda_1 - \lambda_2)R} \left(2\lambda_1 \lambda_2 R e^{2\lambda_2 R} E_1(2\lambda_2 R) - 2\lambda_1 \lambda_2 R e^{2\lambda_1 R} E_1(2\lambda_1 R)\right) \nonumber \\
    &= -\frac{A\sqrt{B_+ B_-}}{|B|} j \frac{2\lambda_1 \lambda_2}{\lambda_1 - \lambda_2} \left(e^{2\lambda_2 R} E_1(2\lambda_2 R) - e^{2\lambda_1 R} E_1(2\lambda_1 R)\right) \nonumber \\
    &= -\frac{A\sqrt{B_+ B_-}}{|B|} \frac{2j}{\lambda_2^{-1} - \lambda_1^{-1}} \left(e^{2\lambda_2 R} E_1(2\lambda_2 R) - e^{2\lambda_1 R} E_1(2\lambda_1 R)\right) .
    \label{eq:analytic}
\end{align}
Written in terms of $d$ gives
\begin{equation}
    E_j = -\frac{A\sqrt{B_+ B_-}}{|B|} \frac{2j}{\lambda_2^{-1} - \lambda_1^{-1}} \left(e^{\lambda_2 d} E_1(\lambda_2 d) - e^{\lambda_1 d} E_1(\lambda_1 d)\right) .
\end{equation}

Both the analytic result and the series expansions to different orders are plotted in Fig.~\ref{fig:hex-analytic-series}.
It can be observed that the crossing between the curves is an artefact from the terms of alternating signs.
The inclusion of each term makes the curves disagree more with the analytic result, showing that the series is divergent.

\end{widetext}

\bibliography{ref}% Produces the bibliography via BibTeX.

@article{Kane2005,
  title = {Quantum {{Spin}} Hall Effect in Graphene},
  author = {Kane, C. L. and Mele, E. J.},
  journal = {Physical Review Letters},
  volume = {95},
  number = {22},
  pages = {226801},
  year = {2005},
  month = nov,
  publisher = {American Physical Society},
  url = {https://journals.aps.org/prl/abstract/10.1103/PhysRevLett.95.226801},
  doi = {10.1103/PhysRevLett.95.226801},
  issn = {00319007},
  pmid = {16384250}
}

@article{Kane2005-Z2,
  title = {${Z}_{2}$ Topological Order and the Quantum Spin Hall Effect},
  author = {Kane, C. L. and Mele, E. J.},
  journal = {Physical Review Letters},
  volume = {95},
  number = {14},
  pages = {146802},
  year = {2005},
  month = sep,
  publisher = {American Physical Society},
  url = {https://link.aps.org/doi/10.1103/PhysRevLett.95.146802},
  doi = {10.1103/PhysRevLett.95.146802}
}

@article{BHZ,
  title = {Quantum Spin Hall Effect and Topological Phase Transition in {{HgTe}} Quantum Wells},
  author = {Bernevig, B. Andrei and Hughes, Taylor L. and Zhang, Shou Cheng},
  journal = {Science},
  volume = {314},
  number = {5806},
  pages = {1757--1761},
  year = {2006},
  month = dec,
  publisher = {American Association for the Advancement of Science},
  url = {https://www.science.org/doi/10.1126/science.1133734},
  doi = {10.1126/science.1133734},
  issn = {00368075},
  pmid = {17170299}
}

@article{Liu2008,
  title = {Quantum {{Spin Hall Effect}} in {{Inverted Type-II Semiconductors}}},
  author = {Liu, Chaoxing and Hughes, Taylor L. and Qi, Xiao-Liang and Wang, Kang and Zhang, Shou-Cheng},
  journal = {Physical Review Letters},
  volume = {100},
  number = {23},
  pages = {236601},
  year = 2008,
  month = jun,
  publisher = {American Physical Society},
  url = {https://link.aps.org/doi/10.1103/PhysRevLett.100.236601},
  doi = {10.1103/PhysRevLett.100.236601}
}

@book{Asboth2016,
  title = {A {{Short Course}} on {{Topological Insulators}}},
  author = {Asboth, Janos K. and Oroszlany, Laszlo and Palyi, Andras},
  journal = {Lecture Notes in Physics},
  volume = {919},
  year = 2016,
  publisher = {Springer International Publishing},
  url = {http://arxiv.org/abs/1509.02295},
  isbn = {978-3-319-25607-8}
}

@article{Qi2011,
  title = {Topological Insulators and Superconductors},
  author = {Qi, Xiao Liang and Zhang, Shou Cheng},
  journal = {Reviews of Modern Physics},
  volume = {83},
  number = {4},
  pages = {1057},
  year = {2011},
  month = oct,
  publisher = {American Physical Society},
  url = {https://journals.aps.org/rmp/abstract/10.1103/RevModPhys.83.1057},
  doi = {10.1103/RevModPhys.83.1057},
  issn = {00346861}
}

@article{Zhou2008,
  title = {Finite {{Size Effects}} on {{Helical Edge States}} in a {{Quantum Spin-Hall System}}},
  author = {Zhou, Bin and Lu, Hai-Zhou and Chu, Rui-Lin and Shen, Shun-Qing and Niu, Qian},
  journal = {Physical Review Letters},
  volume = {101},
  number = {24},
  pages = {246807},
  year = 2008,
  month = dec,
  publisher = {American Physical Society},
  url = {https://link.aps.org/doi/10.1103/PhysRevLett.101.246807},
  doi = {10.1103/PhysRevLett.101.246807}
}

@article{Wada2011,
  title = {Localized Edge States in Two-Dimensional Topological Insulators: {{Ultrathin Bi}} Films},
  author = {Wada, M. and Murakami, S. and Freimuth, F. and Bihlmayer, G.},
  journal = {Physical Review B},
  volume = {83},
  number = {12},
  pages = {121310},
  year = 2011,
  month = mar,
  publisher = {American Physical Society},
  url = {https://link.aps.org/doi/10.1103/PhysRevB.83.121310},
  doi = {10.1103/PhysRevB.83.121310},
  shorttitle = {Localized Edge States in Two-Dimensional Topological Insulators}
}

@article{Lunde2012,
  title = {Helical Edge States Coupled to a Spin Bath: {{Current-induced}} Magnetization},
  author = {Lunde, Anders Mathias and Platero, Gloria},
  journal = {Physical Review B},
  volume = {86},
  number = {3},
  pages = {035112},
  year = 2012,
  month = jul,
  publisher = {American Physical Society},
  url = {https://link.aps.org/doi/10.1103/PhysRevB.86.035112},
  doi = {10.1103/PhysRevB.86.035112},
  shorttitle = {Helical Edge States Coupled to a Spin Bath}
}

@article{Konig2008,
  title = {The {{Quantum Spin Hall Effect}}: {{Theory}} and {{Experiment}}},
  author = {K{\"o}nig, Markus and Buhmann, Hartmut and W. Molenkamp, Laurens and Hughes, Taylor and Liu, Chao-Xing and Qi, Xiao-Liang and Zhang, Shou-Cheng},
  journal = {Journal of the Physical Society of Japan},
  volume = {77},
  number = {3},
  pages = {031007},
  year = 2008,
  month = mar,
  publisher = {The Physical Society of Japan},
  url = {https://journals.jps.jp/doi/10.1143/JPSJ.77.031007},
  doi = {10.1143/JPSJ.77.031007},
  shorttitle = {The {{Quantum Spin Hall Effect}}},
  issn = {0031-9015}
}

@article{Vaitkus2022,
  title = {Effect of Magnetic Impurity Scattering on Transport in Topological Insulators},
  author = {Vaitkus, Jesse A. and Ho, Cong Son and Cole, Jared H.},
  journal = {Physical Review B},
  volume = {106},
  number = {11},
  pages = {115420},
  year = 2022,
  month = sep,
  publisher = {American Physical Society},
  url = {https://link.aps.org/doi/10.1103/PhysRevB.106.115420},
  doi = {10.1103/PhysRevB.106.115420}
}

@article{Weber2024,
  title = {2024 Roadmap on {{2D}} Topological Insulators},
  author = {Weber, Bent and Fuhrer, Michael S. and Sheng, Xian-Lei and Yang, Shengyuan A. and Thomale, Ronny and Shamim, Saquib and Molenkamp, Laurens W. and Cobden, David and Pesin, Dmytro and Zandvliet, Harold J. W. and Bampoulis, Pantelis and Claessen, Ralph and Menges, Fabian R. and Gooth, Johannes and Felser, Claudia and Shekhar, Chandra and Tadich, Anton and Zhao, Mengting and Edmonds, Mark T. and Jia, Junxiang and Bieniek, Maciej and V{\"a}yrynen, Jukka I. and Culcer, Dimitrie and Muralidharan, Bhaskaran and Nadeem, Muhammad},
  journal = {Journal of Physics: Materials},
  volume = {7},
  number = {2},
  pages = {022501},
  year = {2024},
  month = mar,
  publisher = {IOP Publishing},
  url = {https://iopscience.iop.org/article/10.1088/2515-7639/ad2083},
  doi = {10.1088/2515-7639/ad2083},
  issn = {2515-7639}
}

@article{Tretiakov2011,
  title = {Holey Topological Thermoelectrics},
  author = {Tretiakov, O. A. and Abanov, {\relax Ar}. and Sinova, Jairo},
  journal = {Applied Physics Letters},
  volume = {99},
  number = {11},
  pages = {113110},
  year = {2011},
  month = sep,
  url = {https://doi.org/10.1063/1.3637055},
  doi = {10.1063/1.3637055},
  issn = {0003-6951}
}

@article{Fu2018,
  title = {New Topological States in {{HgTe}} Quantum Wells from Defect Patterning},
  author = {Fu, Hua-Hua and Wu, Ruqian},
  journal = {Nanoscale},
  volume = {10},
  number = {33},
  pages = {15462--15467},
  year = 2018,
  month = aug,
  publisher = {The Royal Society of Chemistry},
  url = {https://pubs.rsc.org/en/content/articlelanding/2018/nr/c8nr04878a},
  doi = {10.1039/C8NR04878A},
  issn = {2040-3372}
}

@article{Kariyado2018,
  title = {Counterpropagating Topological Interface States in Graphene Patchwork Structures with Regular Arrays of Nanoholes},
  author = {Kariyado, Toshikaze and Jiang, Yong-Cheng and Yang, Hongxin and Hu, Xiao},
  journal = {Physical Review B},
  volume = {98},
  number = {19},
  pages = {195416},
  year = 2018,
  month = nov,
  publisher = {American Physical Society},
  url = {https://link.aps.org/doi/10.1103/PhysRevB.98.195416},
  doi = {10.1103/PhysRevB.98.195416}
}

@article{Jiang2026,
  title = {Periodic {{Behavior}} of {{Topology}} in {{Graphene}} with {{Nanohole Array}}},
  author = {Jiang, Yong-Cheng and Wang, Xing-Xiang and Hu, Xiao},
  journal = {Journal of the Physical Society of Japan},
  volume = {95},
  number = {6},
  pages = {063707},
  year = 2026,
  month = jun,
  publisher = {The Physical Society of Japan},
  url = {https://journals.jps.jp/doi/10.7566/JPSJ.95.063707},
  doi = {10.7566/JPSJ.95.063707},
  issn = {0031-9015}
}

@article{Geuzaine2009,
  title = {Gmsh: {{A}} 3-{{D}} Finite Element Mesh Generator with Built-in Pre- and Post-Processing Facilities},
  author = {Geuzaine, Christophe and Remacle, Jean-Fran{\c c}ois},
  journal = {International Journal for Numerical Methods in Engineering},
  volume = {79},
  number = {11},
  pages = {1309--1331},
  year = {2009},
  url = {https://onlinelibrary.wiley.com/doi/abs/10.1002/nme.2579},
  doi = {10.1002/nme.2579},
  shorttitle = {Gmsh},
  issn = {1097-0207},
  copyright = {Copyright {\copyright} 2009 John Wiley \& Sons, Ltd.}
}

@misc{Baratta2023,
  title = {{{DOLFINx}}: {{The}} next Generation {{FEniCS}} Problem Solving Environment},
  author = {Baratta, Igor A. and Dean, Joseph P. and Dokken, J{\o}rgen S. and Habera, Michal and Hale, Jack S. and Richardson, Chris N. and Rognes, Marie E. and Scroggs, Matthew W. and Sime, Nathan and Wells, Garth N.},
  year = 2023,
  month = dec,
  publisher = {Zenodo},
  archiveprefix = {Zenodo},
  url = {https://zenodo.org/records/10447666},
  doi = {10.5281/zenodo.10447666},
  shorttitle = {{{DOLFINx}}}
}

@article{Beugeling2025,
  title = {Kdotpy: K{$\cdot$}p Theory on a Lattice for Simulating Semiconductor Band Structures},
  author = {Beugeling, Wouter and Bayer, Florian and Berger, Christian and B{\"o}ttcher, Jan and Bovkun, Leonid and Fuchs, Christopher and Hofer, Maximilian and Shamim, Saquib and Siebert, Moritz and Wang, Li-Xian and Hankiewicz, Ewelina and Kie{\ss}ling, Tobias and Buhmann, Hartmut and Molenkamp, Laurens},
  journal = {SciPost Physics Codebases},
  pages = {047},
  year = {2025},
  month = jan,
  url = {https://scipost.org/SciPostPhysCodeb.47},
  doi = {10.21468/SciPostPhysCodeb.47},
  shorttitle = {Kdotpy},
  issn = {2949-804X}
}

@article{Thouless1982,
  title = {Quantized {{Hall Conductance}} in a {{Two-Dimensional Periodic Potential}}},
  author = {Thouless, D. J. and Kohmoto, M. and Nightingale, M. P. and {den Nijs}, M.},
  journal = {Physical Review Letters},
  volume = {49},
  number = {6},
  pages = {405--408},
  year = {1982},
  month = aug,
  publisher = {American Physical Society},
  url = {https://link.aps.org/doi/10.1103/PhysRevLett.49.405},
  doi = {10.1103/PhysRevLett.49.405}
}

@article{Avron1983,
  title = {Homotopy and {{Quantization}} in {{Condensed Matter Physics}}},
  author = {Avron, J. E. and Seiler, R. and Simon, B.},
  journal = {Physical Review Letters},
  volume = {51},
  number = {1},
  pages = {51--53},
  year = 1983,
  month = jul,
  publisher = {American Physical Society},
  url = {https://link.aps.org/doi/10.1103/PhysRevLett.51.51},
  doi = {10.1103/PhysRevLett.51.51}
}

@article{Simon1983,
  title = {Holonomy, the {{Quantum Adiabatic Theorem}}, and {{Berry}}'s {{Phase}}},
  author = {Simon, Barry},
  journal = {Physical Review Letters},
  volume = {51},
  number = {24},
  pages = {2167--2170},
  year = 1983,
  month = dec,
  publisher = {American Physical Society},
  url = {https://link.aps.org/doi/10.1103/PhysRevLett.51.2167},
  doi = {10.1103/PhysRevLett.51.2167}
}

@article{Niu1985,
  title = {Quantized {{Hall}} Conductance as a Topological Invariant},
  author = {Niu, Qian and Thouless, D. J. and Wu, Yong-Shi},
  journal = {Physical Review B},
  volume = {31},
  number = {6},
  pages = {3372--3377},
  year = {1985},
  month = mar,
  publisher = {American Physical Society},
  url = {https://link.aps.org/doi/10.1103/PhysRevB.31.3372},
  doi = {10.1103/PhysRevB.31.3372}
}

@article{Fukui2005,
  title = {Chern {{Numbers}} in {{Discretized Brillouin Zone}}: {{Efficient Method}} of {{Computing}} ({{Spin}}) {{Hall Conductances}}},
  author = {Fukui, Takahiro and Hatsugai, Yasuhiro and Suzuki, Hiroshi},
  journal = {Journal of the Physical Society of Japan},
  volume = {74},
  number = {6},
  pages = {1674--1677},
  year = {2005},
  month = jun,
  publisher = {The Physical Society of Japan},
  url = {https://journals.jps.jp/doi/10.1143/JPSJ.74.1674},
  doi = {10.1143/JPSJ.74.1674},
  shorttitle = {Chern {{Numbers}} in {{Discretized Brillouin Zone}}},
  issn = {0031-9015}
}

@article{Prodan2009,
  title = {Robustness of the Spin-{{Chern}} Number},
  author = {Prodan, Emil},
  journal = {Physical Review B},
  volume = {80},
  number = {12},
  pages = {125327},
  year = {2009},
  month = sep,
  publisher = {American Physical Society},
  url = {https://link.aps.org/doi/10.1103/PhysRevB.80.125327},
  doi = {10.1103/PhysRevB.80.125327}
}

@article{Esaki1970,
  title = {Superlattice and {{Negative Differential Conductivity}} in {{Semiconductors}}},
  author = {Esaki, L. and Tsu, R.},
  journal = {IBM Journal of Research and Development},
  volume = {14},
  number = {1},
  pages = {61--65},
  year = {1970},
  month = jan,
  url = {https://ieeexplore.ieee.org/document/5391729},
  doi = {10.1147/rd.141.0061},
  issn = {0018-8646}
}

@article{Wallace1947,
  title = {The {{Band Theory}} of {{Graphite}}},
  author = {Wallace, P. R.},
  journal = {Physical Review},
  volume = {71},
  number = {9},
  pages = {622--634},
  year = {1947},
  month = may,
  publisher = {American Physical Society},
  url = {https://link.aps.org/doi/10.1103/PhysRev.71.622},
  doi = {10.1103/PhysRev.71.622}
}

@article{Semenoff1984,
  title = {Condensed-{{Matter Simulation}} of a {{Three-Dimensional Anomaly}}},
  author = {Semenoff, Gordon W.},
  journal = {Physical Review Letters},
  volume = {53},
  number = {26},
  pages = {2449--2452},
  year = {1984},
  month = dec,
  publisher = {American Physical Society},
  url = {https://link.aps.org/doi/10.1103/PhysRevLett.53.2449},
  doi = {10.1103/PhysRevLett.53.2449}
}

@article{DiVincenzo1984,
  title = {Self-Consistent Effective-Mass Theory for Intralayer Screening in Graphite Intercalation Compounds},
  author = {DiVincenzo, D. P. and Mele, E. J.},
  journal = {Physical Review B},
  volume = {29},
  number = {4},
  pages = {1685--1694},
  year = {1984},
  month = feb,
  publisher = {American Physical Society},
  url = {https://link.aps.org/doi/10.1103/PhysRevB.29.1685},
  doi = {10.1103/PhysRevB.29.1685}
}

@article{Novoselov2004,
  title = {Electric {{Field Effect}} in {{Atomically Thin Carbon Films}}},
  author = {Novoselov, K. S. and Geim, A. K. and Morozov, S. V. and Jiang, D. and Zhang, Y. and Dubonos, S. V. and Grigorieva, I. V. and Firsov, A. A.},
  journal = {Science},
  volume = {306},
  number = {5696},
  pages = {666--669},
  year = {2004},
  month = oct,
  publisher = {American Association for the Advancement of Science},
  url = {https://www.science.org/doi/10.1126/science.1102896},
  doi = {10.1126/science.1102896}
}

@article{Novoselov2012,
  title = {A Roadmap for Graphene},
  author = {Novoselov, K. S. and Fal{$\prime$}ko, V. I. and Colombo, L. and Gellert, P. R. and Schwab, M. G. and Kim, K.},
  journal = {Nature},
  volume = {490},
  number = {7419},
  pages = {192--200},
  year = {2012},
  month = oct,
  publisher = {Nature Publishing Group},
  url = {https://www.nature.com/articles/nature11458},
  doi = {10.1038/nature11458},
  issn = {1476-4687},
  copyright = {2012 Springer Nature Limited}
}

@article{Polini2013,
  title = {Artificial Honeycomb Lattices for Electrons, Atoms and Photons},
  author = {Polini, Marco and Guinea, Francisco and Lewenstein, Maciej and Manoharan, Hari C. and Pellegrini, Vittorio},
  journal = {Nature Nanotechnology},
  volume = {8},
  number = {9},
  pages = {625--633},
  year = {2013},
  month = sep,
  publisher = {Nature Publishing Group},
  url = {https://www.nature.com/articles/nnano.2013.161},
  doi = {10.1038/nnano.2013.161},
  issn = {1748-3395},
  copyright = {2013 Springer Nature Limited}
}

@article{Cao2018super,
  title = {Unconventional Superconductivity in Magic-Angle Graphene Superlattices},
  author = {Cao, Yuan and Fatemi, Valla and Fang, Shiang and Watanabe, Kenji and Taniguchi, Takashi and Kaxiras, Efthimios and {Jarillo-Herrero}, Pablo},
  journal = {Nature},
  volume = {556},
  number = {7699},
  pages = {43--50},
  year = {2018},
  month = apr,
  publisher = {Nature Publishing Group},
  url = {https://www.nature.com/articles/nature26160},
  doi = {10.1038/nature26160},
  issn = {1476-4687},
  copyright = {2018 Macmillan Publishers Limited, part of Springer Nature. All rights reserved.}
}

@article{Cao2018insul,
  title = {Correlated Insulator Behaviour at Half-Filling in Magic-Angle Graphene Superlattices},
  author = {Cao, Yuan and Fatemi, Valla and Demir, Ahmet and Fang, Shiang and Tomarken, Spencer L. and Luo, Jason Y. and {Sanchez-Yamagishi}, Javier D. and Watanabe, Kenji and Taniguchi, Takashi and Kaxiras, Efthimios and Ashoori, Ray C. and {Jarillo-Herrero}, Pablo},
  journal = {Nature},
  volume = {556},
  number = {7699},
  pages = {80--84},
  year = {2018},
  month = apr,
  publisher = {Nature Publishing Group},
  url = {https://www.nature.com/articles/nature26154},
  doi = {10.1038/nature26154},
  issn = {1476-4687},
  copyright = {2018 Macmillan Publishers Limited, part of Springer Nature. All rights reserved.}
}

@article{Du2018,
  title = {Emerging Many-Body Effects in Semiconductor Artificial Graphene with Low Disorder},
  author = {Du, Lingjie and Wang, Sheng and Scarabelli, Diego and Pfeiffer, Loren N. and West, Ken W. and Fallahi, Saeed and Gardner, Geoff C. and Manfra, Michael J. and Pellegrini, Vittorio and Wind, Shalom J. and Pinczuk, Aron},
  journal = {Nature Communications},
  volume = {9},
  number = {1},
  pages = {3299},
  year = {2018},
  month = aug,
  publisher = {Nature Publishing Group},
  url = {https://www.nature.com/articles/s41467-018-05775-4},
  doi = {10.1038/s41467-018-05775-4},
  issn = {2041-1723},
  copyright = {2018 The Author(s)}
}

@article{Krix2020,
  title = {Artificial Graphene in a Strong Magnetic Field: {{Bulk}} Current Distribution and Quantum Phase Transitions},
  author = {Krix, Z. E. and Sushkov, O. P.},
  journal = {Physical Review B},
  volume = {101},
  number = {24},
  pages = {245311},
  year = {2020},
  month = jun,
  publisher = {American Physical Society},
  url = {https://link.aps.org/doi/10.1103/PhysRevB.101.245311},
  doi = {10.1103/PhysRevB.101.245311},
  shorttitle = {Artificial Graphene in a Strong Magnetic Field}
}

@article{Krix2022,
  title = {Correlated Physics in an Artificial Triangular Anti-Dot Lattice},
  author = {Krix, Z. E. and Scammell, H. D. and Sushkov, O. P.},
  journal = {Physical Review B},
  volume = {105},
  number = {7},
  pages = {075120},
  year = {2022},
  month = feb,
  publisher = {American Physical Society},
  url = {https://link.aps.org/doi/10.1103/PhysRevB.105.075120},
  doi = {10.1103/PhysRevB.105.075120}
}

@article{Wang2023,
  title = {Formation of {{Artificial Fermi Surfaces}} with a {{Triangular Superlattice}} on a {{Conventional Two-Dimensional Electron Gas}}},
  author = {Wang, Daisy Q. and Krix, Zeb and Sushkov, Oleg P. and Farrer, Ian and Ritchie, David A. and Hamilton, Alexander R. and Klochan, Oleh},
  journal = {Nano Letters},
  volume = {23},
  number = {5},
  pages = {1705--1710},
  year = {2023},
  month = mar,
  publisher = {American Chemical Society},
  url = {https://pubs.acs.org/doi/10.1021/acs.nanolett.2c04358},
  doi = {10.1021/acs.nanolett.2c04358},
  issn = {1530-6984}
}

@article{Wang2026,
  title = {Correlated Insulator in the Kagome Flat Band of a Two-Dimensional Electrostatic Crystal},
  author = {Wang, Daisy Q. and Krix, Zeb and Tkachenko, Olga A. and Tkachenko, Vitaly A. and Chen, Chong and Farrer, Ian and Ritchie, David A. and Sushkov, Oleg P. and Hamilton, Alexander R. and Klochan, Oleh},
  journal = {Nature Physics},
  volume = {22},
  number = {7},
  pages = {1079--1086},
  year = 2026,
  month = jul,
  publisher = {Nature Publishing Group},
  url = {https://www.nature.com/articles/s41567-026-03291-7},
  doi = {10.1038/s41567-026-03291-7},
  issn = {1745-2481},
  copyright = {2026 The Author(s)}
}

@misc{Cioni2026,
  title = {Simulating the {{Haldane}} Model in Ultra-Clean {{GaAs}} Heterostructures},
  author = {Cioni, Francesco and Cavicchi, Lorenzo and Taddei, Fabio and Polini, Marco},
  number = {arXiv:2606.24670},
  year = 2026,
  month = jun,
  publisher = {arXiv},
  eprint = {2606.24670},
  archiveprefix = {arXiv},
  primaryclass = {cond-mat.mes-hall},
  url = {https://arxiv.org/abs/2606.24670},
  doi = {10.48550/arXiv.2606.24670}
}

@article{Chen2021,
  title = {Artificial {{Graphene}} on {{Si Substrates}}: {{Fabrication}} and {{Transport Characteristics}}},
  author = {Chen, Peizong and Zhang, Ningning and Peng, Kun and Zhang, Lijian and Yan, Jia and Jiang, Zuimin and Zhong, Zhenyang},
  journal = {ACS Nano},
  volume = {15},
  number = {8},
  pages = {13703--13711},
  year = {2021},
  month = aug,
  publisher = {American Chemical Society},
  url = {https://doi.org/10.1021/acsnano.1c04995},
  doi = {10.1021/acsnano.1c04995},
  shorttitle = {Artificial {{Graphene}} on {{Si Substrates}}},
  issn = {1936-0851}
}

@article{Wang2013,
  title = {Organic Topological Insulators in Organometallic Lattices},
  author = {Wang, Z. F. and Liu, Zheng and Liu, Feng},
  journal = {Nature Communications},
  volume = {4},
  number = {1},
  pages = {1471},
  year = {2013},
  month = feb,
  publisher = {Nature Publishing Group},
  url = {https://www.nature.com/articles/ncomms2451},
  doi = {10.1038/ncomms2451},
  issn = {2041-1723},
  copyright = {2013 Springer Nature Limited}
}

@article{Shelby2001,
  title = {Experimental {{Verification}} of a {{Negative Index}} of {{Refraction}}},
  author = {Shelby, R. A. and Smith, D. R. and Schultz, S.},
  journal = {Science},
  volume = {292},
  number = {5514},
  pages = {77--79},
  year = {2001},
  month = apr,
  publisher = {American Association for the Advancement of Science},
  url = {https://www.science.org/doi/10.1126/science.1058847},
  doi = {10.1126/science.1058847}
}

@article{Pendry2004,
  title = {Negative Refraction},
  author = {Pendry, {\relax JB}},
  journal = {Contemporary Physics},
  volume = {45},
  number = {3},
  pages = {191--202},
  year = {2004},
  month = may,
  publisher = {Taylor \& Francis},
  url = {https://doi.org/10.1080/00107510410001667434},
  doi = {10.1080/00107510410001667434},
  issn = {0010-7514}
}

@article{Song2018,
  title = {Electron Quantum Metamaterials in van Der {{Waals}} Heterostructures},
  author = {Song, Justin C. W. and Gabor, Nathaniel M.},
  journal = {Nature Nanotechnology},
  volume = {13},
  number = {11},
  pages = {986--993},
  year = {2018},
  month = nov,
  publisher = {Nature Publishing Group},
  url = {https://www.nature.com/articles/s41565-018-0294-9},
  doi = {10.1038/s41565-018-0294-9},
  issn = {1748-3395},
  copyright = {2018 Springer Nature Limited}
}

@article{Sheng2006,
  title = {Quantum {{Spin-Hall Effect}} and {{Topologically Invariant Chern Numbers}}},
  author = {Sheng, D. N. and Weng, Z. Y. and Sheng, L. and Haldane, F. D. M.},
  journal = {Physical Review Letters},
  volume = {97},
  number = {3},
  pages = {036808},
  year = {2006},
  month = jul,
  publisher = {American Physical Society},
  url = {https://link.aps.org/doi/10.1103/PhysRevLett.97.036808},
  doi = {10.1103/PhysRevLett.97.036808}
}

@article{Hung2014,
  title = {Interaction Effects on Topological Phase Transitions via Numerically Exact Quantum {{Monte Carlo}} Calculations},
  author = {Hung, Hsiang-Hsuan and Chua, Victor and Wang, Lei and Fiete, Gregory A.},
  journal = {Physical Review B},
  volume = {89},
  number = {23},
  pages = {235104},
  year = {2014},
  month = jun,
  publisher = {American Physical Society},
  url = {https://link.aps.org/doi/10.1103/PhysRevB.89.235104},
  doi = {10.1103/PhysRevB.89.235104}
}

@article{Wang2020,
  title = {Universal Numerical Calculation Method for the {{Berry}} Curvature and {{Chern}} Numbers of Typical Topological Photonic Crystals},
  author = {Wang, Chenyang and Zhang, Hongyu and Yuan, Hongyi and Zhong, Jinrui and Lu, Cuicui},
  journal = {Frontiers of Optoelectronics},
  volume = {13},
  number = {1},
  pages = {73--88},
  year = {2020},
  month = mar,
  url = {https://doi.org/10.1007/s12200-019-0963-9},
  doi = {10.1007/s12200-019-0963-9},
  issn = {2095-2767}
}

@article{Yu2021,
  title = {Neuron-{{Inspired Steiner Tree Networks}} for {{3D Low-Density Metastructures}}},
  author = {Yu, Haoyi and Zhang, Qiming and Cumming, Benjamin P. and Goi, Elena and Cole, Jared H. and Luan, Haitao and Chen, Xi and Gu, Min},
  journal = {Advanced Science},
  volume = {8},
  number = {19},
  pages = {2100141},
  year = 2021,
  url = {https://onlinelibrary.wiley.com/doi/abs/10.1002/advs.202100141},
  doi = {10.1002/advs.202100141},
  issn = {2198-3844}
}

@article{Maier2017,
  title = {Ballistic Geometric Resistance Resonances in a Single Surface of a Topological Insulator},
  author = {Maier, Hubert and Ziegler, Johannes and Fischer, Ralf and Kozlov, Dmitriy and Kvon, Ze Don and Mikhailov, Nikolay and Dvoretsky, Sergey A. and Weiss, Dieter},
  journal = {Nature Communications},
  volume = {8},
  number = {1},
  pages = {2023},
  year = 2017,
  month = dec,
  publisher = {Nature Publishing Group},
  url = {https://www.nature.com/articles/s41467-017-01684-0},
  doi = {10.1038/s41467-017-01684-0},
  issn = {2041-1723},
  copyright = {2017 The Author(s)}
}

@article{Niyazov2023,
  title = {Effective {{Hamiltonian}} of {{Topologically Protected Qubit}} in a {{Helical Crystal}}},
  author = {Niyazov, R. A. and Aristov, D. N. and Kachorovskii, V. {\relax Yu}.},
  journal = {JETP Letters},
  volume = {118},
  number = {5},
  pages = {376--383},
  year = 2023,
  month = sep,
  url = {https://doi.org/10.1134/S0021364023602361},
  doi = {10.1134/S0021364023602361},
  issn = {1090-6487}
}

@article{Gioia2019,
  title = {Spherical Topological Insulator Nanoparticles: {{Quantum}} Size Effects and Optical Transitions},
  author = {Gioia, L. and Christie, M. G. and Z{\"u}licke, U. and Governale, M. and Sneyd, A. J.},
  journal = {Physical Review B},
  volume = {100},
  number = {20},
  pages = {205417},
  year = 2019,
  month = nov,
  publisher = {American Physical Society},
  url = {https://link.aps.org/doi/10.1103/PhysRevB.100.205417},
  doi = {10.1103/PhysRevB.100.205417},
  shorttitle = {Spherical Topological Insulator Nanoparticles}
}

@article{Governale2020,
  title = {Finite-Size Effects in Cylindrical Topological Insulators},
  author = {Governale, Michele and Bhandari, Bibek and Taddei, Fabio and Imura, Ken-Ichiro and Z{\"u}licke, Ulrich},
  journal = {New Journal of Physics},
  volume = {22},
  number = {6},
  pages = {063042},
  year = 2020,
  month = jun,
  publisher = {IOP Publishing},
  url = {https://dx.doi.org/10.1088/1367-2630/ab90d3},
  doi = {10.1088/1367-2630/ab90d3},
  issn = {1367-2630}
}

@article{Governale2023,
  title = {Topological-Insulator Nanocylinders},
  author = {Governale, Michele and Taddei, Fabio},
  journal = {SciPost Physics Core},
  volume = {6},
  number = {2},
  pages = {032},
  year = 2023,
  month = apr,
  url = {https://www.scipost.org/SciPostPhysCore.6.2.032},
  doi = {10.21468/SciPostPhysCore.6.2.032},
  issn = {2666-9366}
}

@article{Bake2023,
  title = {Top-down Patterning of Topological Surface and Edge States Using a Focused Ion Beam},
  author = {Bake, Abdulhakim and Zhang, Qi and Ho, Cong Son and Causer, Grace L. and Zhao, Weiyao and Yue, Zengji and Nguyen, Alexander and Akhgar, Golrokh and Karel, Julie and Mitchell, David and Pastuovic, Zeljko and Lewis, Roger and Cole, Jared H. and Nancarrow, Mitchell and Valanoor, Nagarajan and Wang, Xiaolin and Cortie, David},
  journal = {Nature Communications},
  volume = {14},
  number = {1},
  pages = {1693},
  year = 2023,
  month = mar,
  publisher = {Nature Publishing Group},
  url = {https://www.nature.com/articles/s41467-023-37102-x},
  doi = {10.1038/s41467-023-37102-x},
  issn = {2041-1723},
  copyright = {2023 Crown}
}
\bibliographystyle{apsrev4-2}

\end{document}